\documentclass[lettersize,journal]{IEEEtran}
\usepackage{amsmath,amsfonts}
\usepackage{algorithmic}
\usepackage{algorithm}
\usepackage{array}
\usepackage[caption=false,font=normalsize,labelfont=sf,textfont=sf]{subfig}
\usepackage{textcomp}
\usepackage{stfloats}
\usepackage{url}
\usepackage{verbatim}
\usepackage{graphicx}
\usepackage{cite}
\usepackage{booktabs}
\usepackage{longtable}
\usepackage{multirow}
\usepackage{array}
\usepackage{siunitx}

\title{Simulating Nighttime Visible Satellite Imagery of Tropical Cyclones Using Conditional Generative Adversarial Networks}

\author{Jinghuai Yao, Puyuan Du, Yucheng Zhao, and Yubo Wang
  \thanks{
This research received no external funding. (Corresponding author: Jinghuai Yao.)

Jinghuai Yao is with the Department of Astronomy, University of Wisconsin–Madison, Madison, WI 53706 USA (e-mail: jyao224@wisc.edu).

Puyuan Du is with the Department of Chemistry and Biochemistry, University of California, Los Angeles, Los Angeles, CA 90095 USA (e-mail: puyuandu@g.ucla.edu).

Yucheng Zhao is with the School of Health Sciences, Guangzhou Xinhua University, Guangzhou, Guangdong 510520, China (yuchengzhao0524@gmail.com).

Yubo Wang is with the Department of Atmospheric and Oceanic Sciences, University of Wisconsin–Madison, Madison, WI 53706 USA (e-mail: xiaoqianwx@outlook.com).
}%
}
\date{March 2024}

\begin{document}

\maketitle

\begin{abstract}
Visible (VIS) imagery is important for monitoring Tropical Cyclones (TCs) but is unavailable at night. This study presents a Conditional Generative Adversarial Networks (CGAN) model to generate nighttime VIS imagery with significantly enhanced accuracy and spatial resolution. Our method offers three key improvements compared to existing models. First, we replaced the L1 loss in the pix2pix framework with the Structural Similarity Index Measure (SSIM) loss, which significantly reduced image blurriness. Second, we selected multispectral infrared (IR) bands as input based on a thorough examination of their spectral properties, providing essential physical information for accurate simulation. Third, we incorporated the direction parameters of the sun and the satellite, which addressed the dependence of VIS images on sunlight directions and enabled a much larger training set from continuous daytime data. The model was trained and validated using data from the Advanced Himawari Imager (AHI) in the daytime, achieving statistical results of SSIM = 0.923 and Root Mean Square Error (RMSE) = 0.0299, which significantly surpasses existing models. We also performed a cross-satellite nighttime model validation using the Day/Night Band (DNB) of the Visible/Infrared Imager Radiometer Suite (VIIRS), which yields outstanding results compared to existing models. Our model is operationally applied to generate accurate VIS imagery with arbitrary virtual sunlight directions, significantly contributing to the nighttime monitoring of various meteorological phenomena.
\end{abstract}

\begin{IEEEkeywords}
	Clouds, tropical cyclone (TC), conditional generative adversarial network (CGAN), deep learning, visible (VIS), nighttime, AHI.
\end{IEEEkeywords}

\section{Introduction}
\IEEEPARstart{S}{atellite} imagery has wide applications in monitoring the earth's ocean and land, the pattern of climate change, and extreme weather. Observational capabilities of geostationary meteorological satellites have greatly improved over the past decades, providing near real-time visible (VIS) and infrared (IR) imagery at high spatiotemporal resolutions. Representative geostationary meteorological satellites at use include Himawari-8/9 \cite{bessho2016introduction}, Geostationary Operational Environmental Satellite (GOES)-16/17/18 \cite{schmit2005introducing}, Geostationary Korea Multi-Purpose Satellite-2 Atmosphere (GK-2A) \cite{kim2021introduction}, and Fengyun-4A/B  \cite{yang2017introducing}. At nadir, they usually have a resolution of 0.5 km for the red visible band and a resolution of 2 km for the infrared bands. They typically have an observation cycle of 10 to 15 minutes for the full-disk imagery of the earth, while some have regional observations covering $\sim$1000 km and provide imagery every 2.5 minutes or less \cite{bessho2016introduction, schmit2005introducing}. Regional observations are commonly used to observe the development of Tropical Cyclones (TCs) and other mesoscale systems.

VIS images indicate the intensity of sunlight reflected by the earth. For the red visible band, the ocean and most of the land appear dark in the image, while the areas with cloud cover appear bright, which makes it especially useful in cloud feature detection. IR images show the brightness temperatures ($T_b$) converted from the thermal emission from the earth. In IR images, most of the land, ocean, and low-level clouds appear dark, while the upper-level clouds appear bright. Therefore, it is relatively difficult to detect low-level cloud features directly using IR images \cite{conway1997introduction}.

VIS images have a variety of unique and important applications in TC monitoring. For example, the curved bands of low-level clouds of TCs could be clearly identified in VIS images, which can be used to determine the position of exposed Low-Level Circulation Center (LLCC) according to the Dvorak Technique \cite{dvorak1984tropical}. Besides, visible light could penetrate the cirrus in the TC eyes, which helps to monitor the eye formation process. On the contrary, the positioning of TCs is rather difficult in IR images because of the low contrast of low-level clouds, and the developing eye of TCs cannot be effectively identified in IR images because of the relatively high imaginary index of refraction at corresponding wavelengths \cite{warren2008optical}.

Due to the unavailability of geostationary VIS images at night, the effectiveness of weather monitoring is limited. There are two main categories of existing approaches to provide nighttime satellite images similar to VIS images. One of them is the Day/Night Band (DNB) on Sun-Synchronous Orbit (SSO) satellites. DNB receives the moonlight reflected by the earth at night, and its images have very similar properties to VIS images. The Visible/Infrared Imager Radiometer Suite (VIIRS) carried by Suomi National Polar-orbiting Partnership (SNPP) and NOAA-20/21 has a DNB with 750m resolution at nadir \cite{liao2013suomi, wang2020evaluation}. However, DNB images do not completely fulfill the need for nighttime weather monitoring because high-quality images are only available at certain lunar phases, and each SSO satellite typically only provides one image for a given location every night.

Another approach adopted by recent studies is to use Deep Learning (DL) models to train IR and VIS images in pairs and generate nighttime VIS images from IR images \cite{kim2019nighttime, kim2020impact, harder2020nightvision, han2022hypothetical, cheng2022creating, yan2023simulation, pasillas2024turning}. In \cite{kim2019nighttime}, a Conditional Generative Adversarial Networks (CGAN) model is applied to generate VIS images from single-band IR images using the pix2pix framework \cite{isola2017image}. Following studies adopting similar CGAN models demonstrate that using more IR bands significantly improves the model performance \cite{kim2020impact, han2022hypothetical}. There are also attempts to incorporate climate reanalysis data in model input \cite{cheng2022creating}. Apart from CGAN, various models including K-nearest neighbor regression, UNet, UNet++ \cite{harder2020nightvision}, Deep Neural Networks (DNN) \cite{yan2023simulation}, and Feed-forward Neural Networks (FNN) \cite{pasillas2024turning} have been applied to generate nighttime VIS images. While most of the existing studies trained their models using daytime image pairs \cite{kim2019nighttime, kim2020impact, harder2020nightvision, han2022hypothetical, cheng2022creating, yan2023simulation}, some studies have adopted VIIRS DNB data to directly train on nighttime image pairs \cite{chirokova2023proxyvis, pasillas2024turning}.

Most of the existing DL models successfully achieved qualitative similarity to VIS images. However, the quantitative performances of current models are suboptimal as demonstrated in Section IV. In addition, most of the existing studies focused on full-disk or large-scale ($\sim$3000 km) simulations, and the generated images are usually resampled to lower resolutions as a trade-off for computational costs. The lack of accuracy and spatial resolution limits the effectiveness of current models in the practical monitoring of weather systems.

This study presents a DL approach that generates nighttime visible imagery of TCs and tropical oceans with greatly improved accuracy and spatial resolution. We adopted the CGAN model because of its versatility in image-to-image translation tasks \cite{alotaibi2020deep} and its robust generalization at night according to \cite{harder2020nightvision}. We made substantial improvements to the loss function and the model input. We replaced the L1 loss in pix2pix with the Structural Similarity Index Measure (SSIM) loss to minimize the blurriness of simulated clouds. For model input, multispectral IR bands are selected based on a thorough examination of their spectral properties, and direction parameters of the sun and the satellite are included because VIS reflectance largely depends on them. In contrast to existing studies which only used images at a fixed time of each day \cite{kim2019nighttime, kim2020impact, han2022hypothetical}, the adoption of direction parameters also allowed us to utilize continuous data over daytime, as the effects of changing sunlight are taken into consideration. This consequently makes a much larger training set available. In addition, we cropped data pairs into reasonably small sizes and used data at the original resolutions of IR bands, which facilitates cloud feature detection at smaller scales.

In the following, Section II introduces the satellite data used and explains the details of improvements of model input. Section III focuses on the model architecture and the methods of model validation. Section IV presents model validations for both daytime and nighttime, as well as model applications in TC monitoring. Section V discusses the strengths and limitations of our current model and outlines future work, while Section VI presents the conclusions.

\section{Data}
\subsection{Satellite data}

This study mainly used L1B data products from Advanced Himawari Imager (AHI) on Himawari-8/9. Himawari-8 entered operational service on July 7, 2015, and Himawari-9 replaced Himawari-8 as the primary satellite to provide operational service on December 13, 2022.

The AHI data has a total of 16 spectral bands. There are 3 VIS bands, 1 Near Infrared (NIR) band, 2 Short Wavelength Infrared (SWIR) bands, 1 Medium Wavelength Infrared (MWIR) band, 3 Water Vapor (WV) bands, and 6 Long Wavelength Infrared (LWIR) bands. The spatial resolution of the red visible band (Band03, \textit{$\lambda$=}645 nm) is 0.5 km at nadir, the 2 other VIS bands and the NIR band (Band01, Band02, Band04) have a resolution of 1 km, and all other bands have a resolution of 2 km. Table I summarizes the detailed characteristics of the 16 bands.

\begin{table}[h]
\begin{center}
\footnotesize
\caption{Characteristics of AHI spectral Bands}
\begin{tabular}{m{0.5cm}<{\centering}m{0.7cm}<{\centering}m{1.2cm}<{\centering}m{1.2cm}<{\centering}m{2.5cm}<{\centering}}
\toprule[1pt]
Band & Type & Wavelength (\textit{$\mu$m}) & Resolution (km) & Applications \\
\midrule
01 & VIS & 0.455 & 1.0 & Vegetation, aerosol\\
02 & VIS & 0.510 & 1.0 & Vegetation, aerosol\\
03 & VIS & 0.645 & 0.5 & Low cloud, fog\\
04 & NIR & 0.860 & 1.0 & Vegetation, aerosol, coastline\\
05 & SWIR & 1.610 & 2.0 & Cloud phase, snow\\
06 & SWIR & 2.260 & 2.0 & Cloud phase, particle size\\
07 & MWIR & 3.850 & 2.0 & Low cloud, fog, desert, fire\\
08 & WV & 6.250 & 2.0 & Upper-level water vapor\\
09 & WV & 6.950 & 2.0 & Mid-level water vapor\\
10 & WV & 7.350 & 2.0 & Lower-level water vapor\\
11 & LWIR & 8.600 & 2.0 & $\rm SO_2$, cloud phase\\
12 & LWIR & 9.630 & 2.0 & Ozone\\
13 & LWIR & 10.45 & 2.0 & Clean IR, sea surface temperature\\
14 & LWIR & 11.20 & 2.0 & Clean IR, sea surface temperature\\
15 & LWIR & 12.35 & 2.0 & Cloud phase, sea surface temperature\\
16 & LWIR & 13.30 & 2.0 & $\rm CO_2$, cloud top height\\
\bottomrule[1pt]
\end{tabular}
\end{center}
\end{table}
This study used both target area data and full-disk data of AHI for model training and validation. We used VIIRS DNB Sensor Data Records (SDR) data to perform additional model validations, and we used Advanced Scatterometer (ASCAT) Coastal Winds data to demonstrate the application of our model in TC monitoring \cite{verhoef2012high}. 

The AHI L1B data were obtained from the National Institute of Information and Communications Technology (NICT) Science Cloud. The VIIRS SDR data were obtained from the Comprehensive Large Array-data Stewardship System (CLASS) of the National Oceanic and Atmospheric Administration (NOAA). The ASCAT Coastal Winds data were obtained from The European Organisation for the Exploitation of Meteorological Satellites (EUMETSAT).

\subsection{Band selection}
We used Band03 to generate the VIS images as the ground truth, and we selected 7 IR bands (Band08, 09, 10, 11, 13, 15, 16), 4 direction parameters (solar zenith angle, solar azimuth angle, satellite zenith angle, and satellite azimuth angle), and a land basemap in the input datasets for prediction.

Band03 is the most commonly used VIS band in TC monitoring because of its relatively low reflectance for land and ocean \cite{kidder1995satellite}. Fig. 1a shows the Band03 image of Typhoon Saola (2023) on September 1.

The reflectance of VIS bands and the brightness temperature ($T_b$) of IR bands are influenced by a variety of cloud properties \cite{minnis1998parameterizations}, among which the cloud top height contributes a large part of the difference between IR and VIS images. The temperature of the troposphere generally decreases with height, causing higher clouds to emit less thermal emission, which results in lower $T_b$. For example, thin cirrus usually have lower $T_b$ than cumulus with minor vertical development. Although cumulus appears brighter in VIS images, it is typically darker than cirrus in IR images.

Fig. 1b shows the difference between VIS (Fig. 1a) and normalized IR images of Typhoon Saola (2023). Clear sea surface and cumulonimbus are colored white and have about the same brightness in both images. The cirrus colored in blue appears brighter in the IR image, and the cumulus colored in red appears brighter in the VIS image. Fig. 1c shows the 2D histogram between the same pair of VIS and IR images. Cloud types with different cloud top heights can be identified by their locations in the diagram \cite{raffaelli1995cloud, seze1987time}.

\begin{figure*}[htbp]
	\centering
	\includegraphics[width=7in]{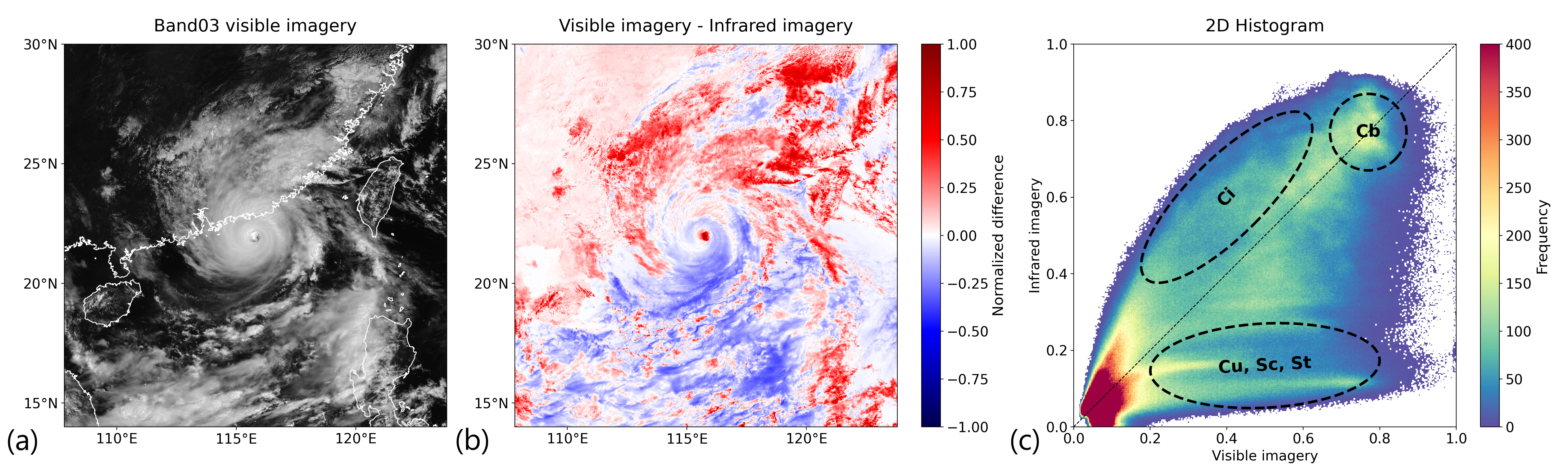}
	\caption{(a) AHI Band03 imagery, (b) difference between normalized VIS and IR imagery, and (c) 2D histogram of Typhoon Saola (2023) in the South China Sea at 03:00 UTC on September 1. The cloud types in (c) are denoted as Cb (Cumulonimbus), Ci (Cirrus),  Cu (Cumulus), Sc (Stratocumulus), and St (Stratus).}
\end{figure*}

While some recent studies used the correlation coefficients between VIS and IR images to find the bands for their model input \cite{kim2020impact, han2022hypothetical}, we chose to find the appropriate IR bands by examining their spectral characteristics and see which bands could derive helpful cloud properties. In order to ``restore" VIS images from IR images, our model should be able to remove the effect of cloud top height on $T_b$. Thus, we selected the combination of IR bands that could help derive cloud top height, so that the DL model can utilize related physical information to remove its effects on IR images. Cloud phase is also indicative of cloud top height as clouds in the upper troposphere mainly consist of ice particles, while clouds in the lower troposphere mainly consist of water droplets \cite{jakel2013thermodynamic}. Thus, cloud phase information could complement the detection of cloud top height, so we also selected the IR bands that could help determine cloud phase. We did not make attempts to directly retrieve these cloud properties quantitatively. Instead, we relied on previous studies on cloud properties retrieval as the theoretical basis to select the input IR bands \cite{menzel1997cloud, menzel2008modis, lutz2003notes}. We used the DL model to perform the derivation of cloud properties implicitly and simulate the VIS images directly from multispectral IR images. The plausibility of this approach is also supported by recent studies that perform cloud property derivation using DL methods \cite{gupta2022cloud, tan2021detecting}.

The underlying assumption of this study is that the selected IR bands exhibit consistent characteristics from day to night. This naturally excludes the NIR and SWIR bands, which are reflective (mainly receiving reflected sunlight), as well as the MWIR band, which is reflective during the day and emissive (mainly receiving the Earth's thermal emission) at night \cite{goody1995atmospheric}.

All WV and LWIR bands are consistently emissive. Among them, we adopted 7 bands that are useful in either determining cloud top height or cloud phase, which are Band08, 09, 10, 11, 13, 15, and 16.

Band13 is considered the ``clean IR" band with the least atmospheric absorption, which makes it representative of cloud patterns \cite{Schmit2018ApplicationsOT}. It is also the most commonly used IR band in TC monitoring. We included Band13 as a reference band to compare with other bands with significant absorption.

Band16 serves the function of determining cloud top height. The most widely adopted approach to derive cloud top height from IR data is the $\rm CO_2$ slicing technique, which relies on the $T_b$ of $\rm CO_2$ absorption bands. The thermal emissions emitted by the clouds with higher cloud top height experience less $\rm CO_2$ absorption, so the $T_b$ difference between the $\rm CO_2$ absorption band and the ``clean IR" band is smaller for higher clouds. For details of the $\rm CO_2$ slicing technique, see \cite{menzel1997cloud, menzel2008modis}. Band16 is the only AHI band with significant $\rm CO_2$ absorption, so we suggest that Band16 would contribute to determining cloud top height, which enhances the performance of our DL model.

The 3 WV bands, Band08--10, complement Band16 to determine cloud top height. While Band16 is dominated by $\rm CO_2$ absorption, Band08--10 has significant $\rm H_2O$ absorption, which could also be utilized to infer cloud top height. This is supported by the MSG/SCE algorithm which includes two WV bands to retrieve cloud top height \cite{lutz2003notes}. In addition, while the peak of the weighting function of Band16 under standard conditions is around 850 hPa, the peak of the weighting function of Band08, Band09, and Band10 are around 350 hPa, 450 hPa, and 550 hPa respectively as shown in Fig. 2. The weighting functions are calculated using the model presented in \cite{saunders2018update}. The peak of the weighting function indicates the level at which a given band has the most absorption and emission \cite{wu2020best}. This corresponds to the optimal level for the band to accurately derive cloud top heights. Therefore, while Band16 serves to determine the height of low-level clouds, the 3 WV bands are more effective for estimating the cloud top height of medium and upper-level clouds.

\begin{figure}[htbp]
	\centering
	\includegraphics[width=3in]{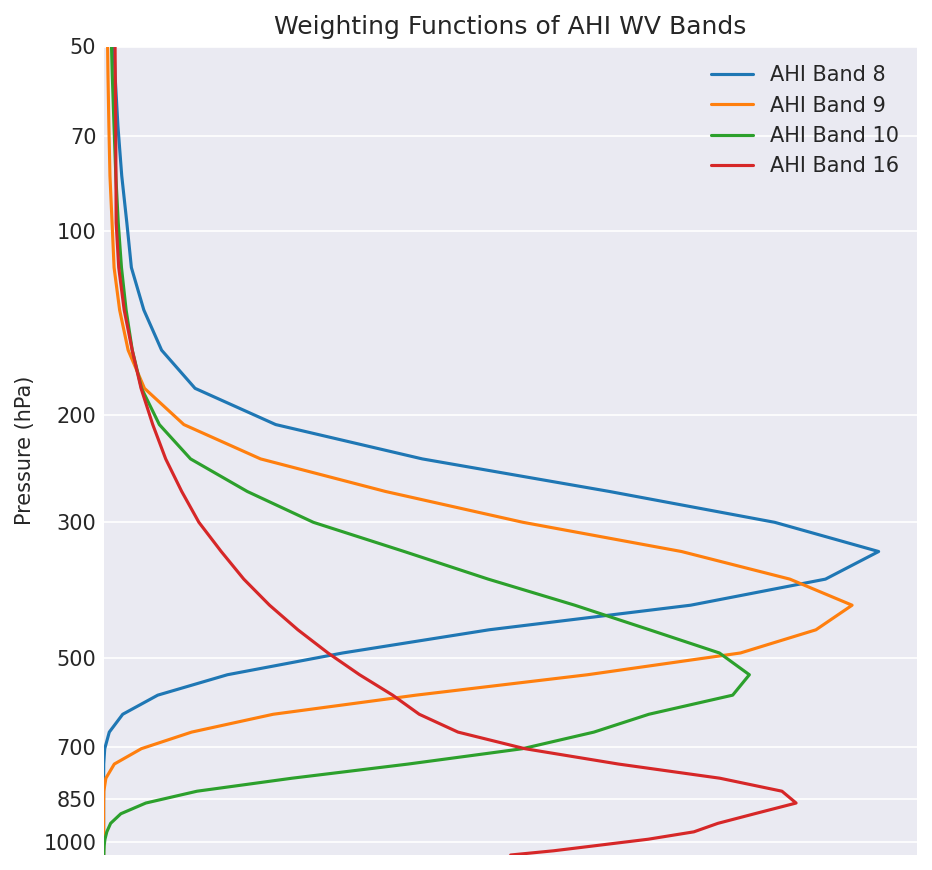}
	\caption{Weighting functions of AHI Band08--10 and Band16 under Standard Tropical Atmosphere and a satellite zenith angle of 0°.}
\end{figure} 

\begin{figure}[htbp]
	\centering
	\includegraphics[width=3.2in]{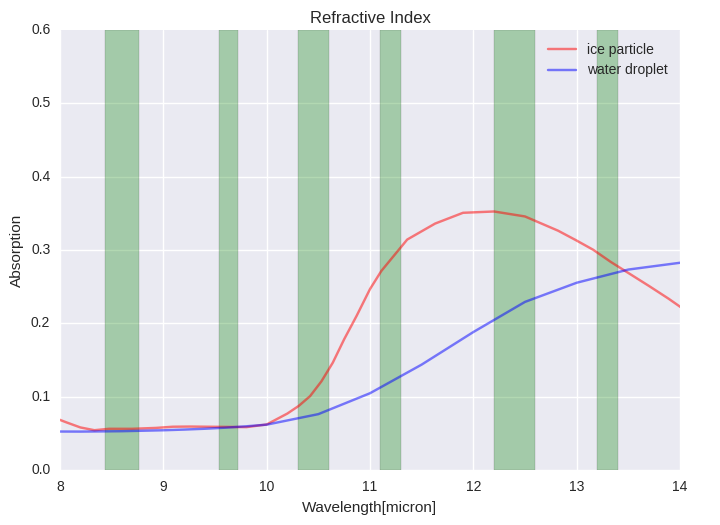}
	\caption{The imaginary index of refraction of water droplets and ice particles and the wavelength range of AHI Band11--16. Source: Meteorological Satellite Center of Japan Meteorological Agency.}
\end{figure} 

Band15 and Band11 serve the function of determining the cloud phase. In particular, these two bands can distinguish thin cirrus, which the IR and VIS images have a big difference on. Current methods to retrieve cloud phase rely on the imaginary index of refraction of cloud particles at different wavelengths \cite{menzel1997cloud}. The index of refraction $\underline{n}$ is given by Eq. (1):

\begin{equation}
\underline{n} = n+i\kappa
\end{equation}

The real portion $n$ indicates the strength of scattering, and the imaginary portion $\kappa$ indicates the strength of absorption. The variation of the imaginary index of refraction of water droplets and ice particles at different infrared wavelengths has been studied by \cite{hale1973optical} and \cite{warren2008optical}. Fig. 3 shows the imaginary index of refraction of water droplets and ice particles over the wavelength range of AHI Band11--16. Among all 6 bands, Band11 has the lowest imaginary index of refraction for ice particles while Band15 has the highest. Consequently, cirrus absorbs more emission at Band15, thus having lower $T_b$ in Band15 than in Band11. We suggest that including Band11 and Band15 contributes to the model performance by identifying the presence of ice particles in clouds, which is also indicative of cloud top height.

The difference between $T_b$ of two bands ($\Delta T_b$) has been widely applied in algorithms of cloud detection and cloud property retrieval \cite{ellrod1995advances, calvert2010goes}. Therefore, we processed the selected bands by calculating $\Delta T_b$ between them, which emphasizes critical information about cloud properties. Details about the $\Delta T_b$ used are covered in Section II.\textit{C}.

We noticed that the VIS reflectance largely depends on the solar zenith angle and the solar azimuth angle. A larger solar zenith angle results in lower overall reflectance and more pronounced lights and shades. Current algorithms of solar zenith angle correction are mainly designed for cloudless areas \cite{richter2005atmospheric}, which only correct the difference in overall reflectance but fail to correct the difference in lights and shades on the cloud top. Apart from that, the solar azimuth angle affects the direction of lights and shades, and we have not found any established algorithm to correct the effect. As it is nearly impossible to fully normalize VIS images to negate the effects of sunlight directions, we included the solar zenith angle and the solar azimuth angle in our input datasets to account for their effects.

Satellite viewing angles also create differences in the correspondence of VIS and IR images, especially for the full-disk data. The satellite viewing angles affect the atmospheric absorption and the projected lengths of lights and shades at the edge of the full-disk observation. Additionally, the satellite viewing angles contribute to the approximate simulation of the sun glare. Therefore, we included the satellite zenith angle and the satellite azimuth angle in our input datasets. We also included a basemap extracted from the red channel of NASA's Blue Marble Next Generation \cite{stockli2005blue} which provides information about land usually not being available in IR bands.

In conclusion, we select a total of 13 channels of model input in our datasets, which includes 1 VIS Band (Band03), 7 IR Bands (Band08, 09, 10, 11, 13, 15, 16), 4 channels of direction parameters (solar zenith angle, solar azimuth angle, satellite zenith angle, and satellite azimuth angle), and 1 channel of basemap.

\subsection{Preprocessing of model input}
Unlike previous studies that used data from a fixed time each day \cite{kim2019nighttime, kim2020impact, han2022hypothetical}, the adoption of sunlight direction parameters allows us to use continuous daytime data. We adopted daytime AHI target area data with TCs as the target from July 7, 2015 to September 6, 2023 at a time interval of 10 minutes and AHI full-disk data once every 15 days from July 7, 2015 to September 22, 2023 at a time interval of 1 hour within the day. Quality controls are performed to exclude the dark data with the central pixel's solar zenith angle $>$ 80° or affected by solar eclipses. After quality control, we obtained a total of 246078 pairs of data, among which 182555 pairs are full-disk data and 63523 pairs are target area data. Our datasets are significantly larger than those used in previous studies.

Target area data has a dimension of 2000×2000 pixels for Band03 and 500×500 pixels for Band05--16. The full-disk data is cropped into neighboring blocks of the same size as the target area data. The boundary of blocks is randomly selected so that overfitting at particular geographical positions can be prevented. Then, the VIS images are resampled to an array of 500×500 pixels. Both VIS and IR images are plotted on a canvas of 512×512 pixels as the model input.

We normalized L1B data into arrays within the range [0, 1] before using them as the model input. For Band03, we clipped the retrieved reflectance in the range of [0, 1] to eliminate the pixels of overexposure. For the basemap, we converted the brightness of the red channel of NASA's Blue Marble Next Generation \cite{stockli2005blue} into values within [0, 1]. For the infrared bands, we normalized $T_b$ and $\Delta T_b$ into the range [0, 1] with certain threshold values $T_{max}$ and $T_{min}$:

\begin{equation}
T_{[0, 1]} = \frac{T_b - T_{min}}{T_{max}-T_{min}}
\end{equation}

Similarly, we also normalized the 4 direction parameters within the range [0, 1] with threshold values:

\begin{equation}
\theta_{[0, 1]} = \frac{\theta - \theta_{min}}{\theta_{max}-\theta_{min}}
\end{equation}

Table II shows detailed information of $T_{max}$, $T_{min}$, $\theta_{max}$, and $\theta_{min}$ for each type of data in the model input. $T_{max}$ and $T_{min}$ are determined based on the common temperature range of $T_b$ and $\Delta T_b$ on full-disk imagery.

\begin{table}[h]
\begin{center}
\footnotesize
\caption{Threshold values applied in the normalization of IR data and direction parameters}
\begin{tabular}{ccccc}
	\toprule
	Channel & $T_{max}$ (°C) & $T_{min}$ (°C) & $\theta_{max}$ (°) & $\theta_{min}$ (°) \\
	\midrule
	Band13 & 45 & -103 & - & -\\
	Band13 - Band08 & 80 & -11 & - & -\\
	Band13 - Band09 & 70 & -10 & - & -\\
	Band13 - Band10 & 62 & -12 & - & -\\
	Band11 - Band15 & 22 & -12 & - & -\\
	Band13 - Band15 & 22 & -3 & - & -\\
	Band13 - Band16 & 41 & -3 & - & -\\
	Zenith angle & - & - & 90 & 0\\
	Azimuth angle & - & - & 180 & -180\\
	\bottomrule
\end{tabular}
\end{center}
\end{table}

The normalized $T_{[0, 1]}$ and $\Delta T_{[0, 1]}$ are generally closer to 0 for cloud-covered regions and closer to 1 for cloudless regions, which means they are inversely correlated with VIS reflectance. To make the training process slightly easier, we inverted the normalized $T_{[0, 1]}$ as follows:

\begin{equation}
T_{[0, 1]}^{'} = 1-T_{[0, 1]}
\end{equation}

Fig. 4 visualizes the preprocessed IR data, direction parameters, basemap, and VIS reflectance of Typhoon Chan-hom (2015) at 05:00 UTC on July 10. For the discussions in Section III, we denote the collection of the 12 channels on the left as $X$ and the VIS reflectance on the right as $Y$.

\begin{figure}[htbp]
	\centering
	\includegraphics[width=3.5in]{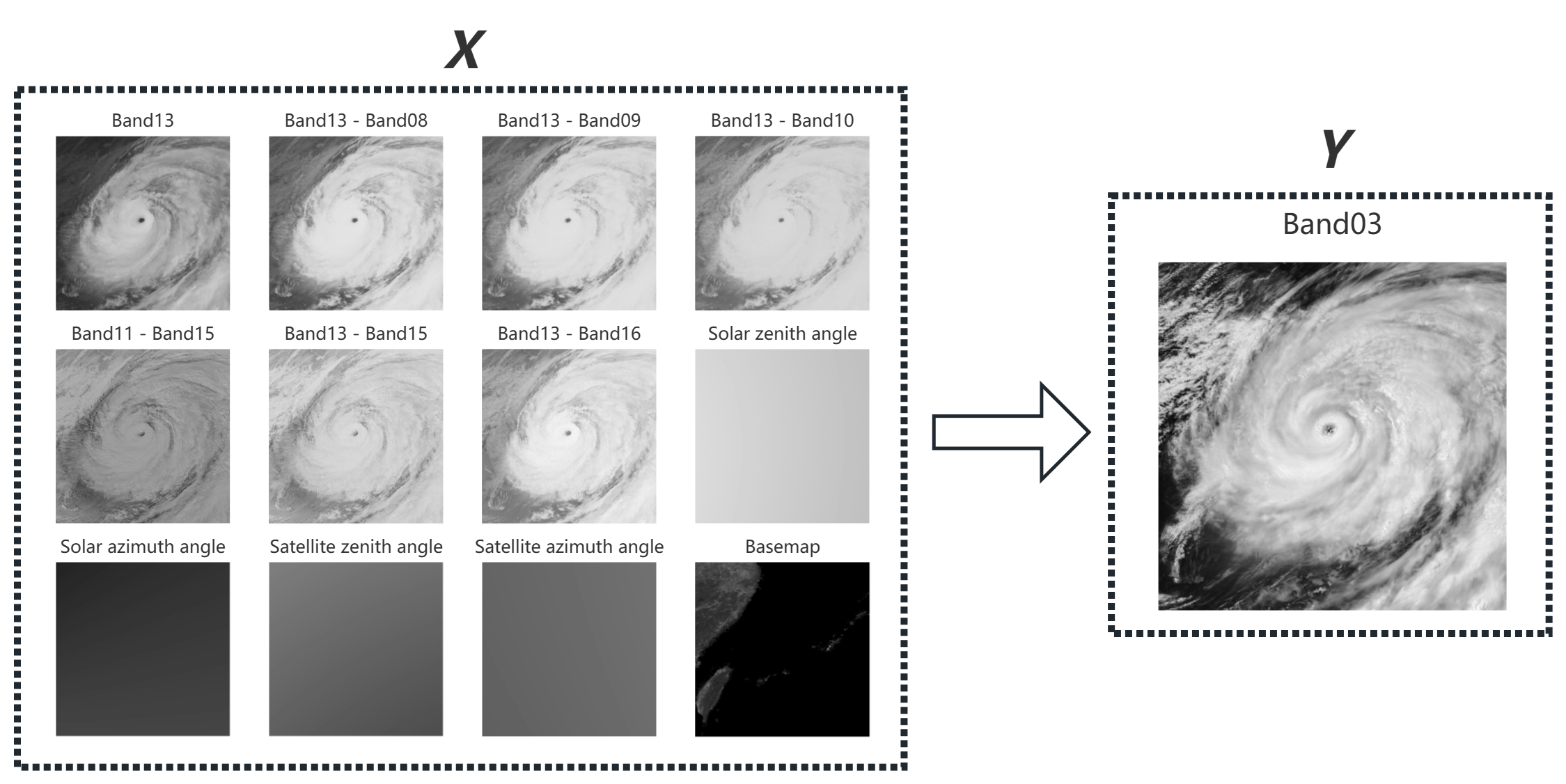}
	\caption{The preprocessed IR data, direction parameters, basemap, and VIS reflectance of Typhoon Chan-hom (2015) at 05:00 UTC on July 10.}
\end{figure} 

\section{Methods}
\subsection{The revised pix2pix model}

Multiple studies \cite{kim2019nighttime, kim2020impact, han2022hypothetical} on simulating nighttime satellite imagery adopted the pix2pix model \cite{isola2017image}, which belongs to the category of CGAN models. Our DL model is also primarily based on the pix2pix model, and we modified its structure and loss function to improve its performance on our datasets. We will first introduce the basic structure of the pix2pix model, followed by a discussion of our modifications.

As GANs, the pix2pix model is composed of two parts: the generator (G) and the discriminator (D). G generates a virtual output $Y^{'}$ from its input $X$ as shown in Fig. 4. G encodes and decodes the input channels with a U-Net structure \cite{ronneberger2015u}, which extracts the underlying features in the input channels at different levels. We suggest this design enables the pix2pix model to extract information about TC structures of different sizes such as the eye, the rainbands, and the central dense overcast (CDO). D discriminates whether an image is a real image or an AI-generated image. The input of D is either paired $\{X, Y\}$ or paired $\{X, Y^{'}\}$, and the output of D is the probability of the image being a real image.

The loss function of the pix2pix model is defined as follows:

\begin{equation}
L_{pix2pix}=\mathop{min}\limits_{G}\mathop{max}\limits_{D}L_{CGAN}(G, D)+k\cdot L_1(G)
\end{equation}

Here, the $\mathop{min}\limits_{G}\mathop{max}\limits_{D}$ notation indicates that $L_{CGAN}(G, D)$ is a function that G aims to minimize and D aims to maximize. $L_{CGAN}(G, D)$ is defined by the following two term function:

\begin{equation}
L_{CGAN}(G, D)=E(log(D(X, Y)))+E(log(1-D(X, Y^{'})))
\end{equation}

The first term $E(log(D(X, Y)))$ evaluates D's performance in discriminating real image as real, while the second term $E(log(1-D(X, Y^{'})))$ evaluates D's performance in discriminating the image generated by G as non-real.

As the other component of $L_{pix2pix}$, the L1 loss $L_1(G)$ is defined as follows:

\begin{equation}
L_1(G)=E(\Vert Y-Y^{'}\Vert)
\end{equation}

$L_1(G)$ ensures that while G is generating $Y^{'}$ that ``looks real" according to D, it is also close enough to the real image $Y$. The coefficient $k$ in Eq. (5) controls the trade-off between $L_1(G)$ and $L_{CGAN}(G, D)$, which is set at $k=100$ by default.

\begin{figure*}[htbp]
	\centering
	\includegraphics[width=6.5in]{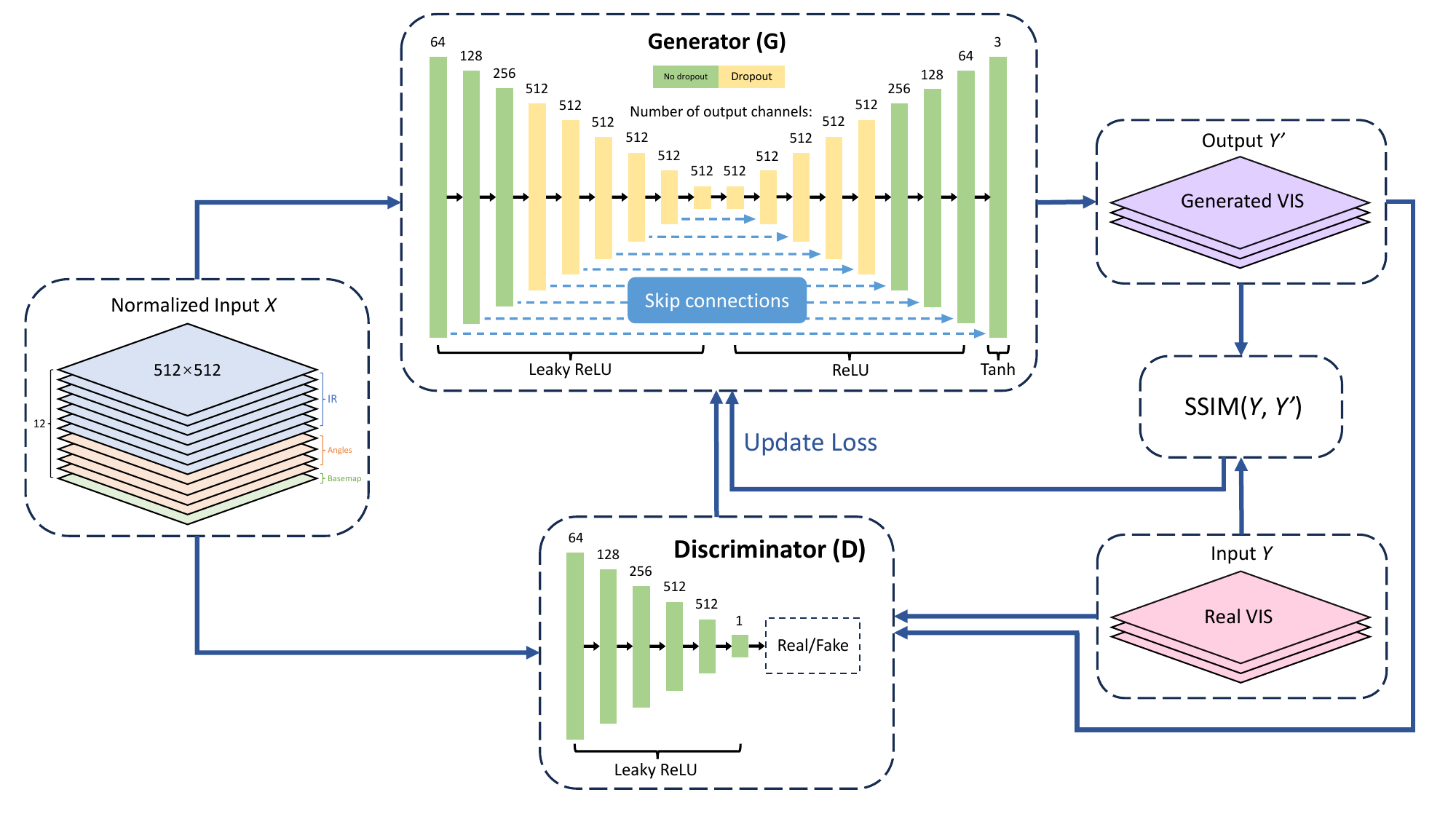}
	\caption{Detailed architecture of the revised pix2pix model. Note that the VIS images are replicated to 3 channels, enabling future framework generalization for multi-color imagery.}
\end{figure*} 

Compared to the original pix2pix model, our model is designed for more complex application scenarios. Therefore, we modified the architectures of both G and D. 

For input $X$ of G, instead of having $256\times256$ images with 3 channels, we have $512\times512$ images with 12 channels. If the original U-Net structure were applied directly, the receptive field would be too small relative to the size of $X$. To expand the receptive field and more comprehensively capture global features, we adopted a deeper U-Net architecture with nine downsampling layers. For activation functions, we use LeakyReLU in the encoder to extract subtle features from the input, ReLU in the first eight layers of the decoder for efficient reconstruction, and Tanh in the final decoding layer for normalization. Additionally, we increased the usage of dropout to mitigate overfitting, especially given the complexity of the IR data. Specifically, dropout is introduced in eleven deep layers of U-Net. 

The input of D concatenates $X$ with either $Y$ or $Y'$, forming $512\times512$ images with 15 channels. Note that Y is replicated to 3 channels, which allows framework generalization for multi-color imagery (e.g. true-color imagery). With a similar consideration of receptive fields, we use a deeper network with six convolutional layers, each employing LeakyReLU as the activation function. Fig. 5 illustrates the details of the revised model architecture.

The pix2pix model used the L1 loss with the reasoning that it would cause less blurriness than the L2 reconstruction loss \cite{isola2017image}. However, when training our model with the L1 loss, we still found certain blurriness on the cloud top. The blurriness or the loss of structural information in the images is particularly significant when the solar zenith angle is larger than $30^\circ$ and apparent lights and shades of clouds exist. We argue that this is because the L1 loss prohibited the generation of lights and shades as simulating them could increase the L1 loss at first. Thus, the model would ``smooth" the lights and shades to minimize the L1 loss, causing blurriness.

Inspired by \cite{harder2020nightvision}, we replaced the L1 loss with the Structural Similarity Index Measure (SSIM) \cite{wang2004image} loss to address this problem. SSIM is a comprehensive measure that considers the similarity between two pictures in luminance, contrast, and structure and is calculated as follows:

\begin{equation}
SSIM(x,y)=\frac{(2\mu_x\mu_y+c_1)(2\sigma_{xy}+c_2)}{(\mu_x^2+\mu_y^2+c_1)(\sigma_x^2+\sigma_y^2+c_2)}
\end{equation}

Here, $\mu_x$ and $\mu_y$ refer to the mean values of the two pictures, $\sigma_x$ and $\sigma_y$ refer to their variances, and $\sigma_{xy}$ refers to their covariance. $c_1$ and $c_2$ have values of $(0.01\cdot L)^2$ and $(0.03\cdot L)^2$ respectively, where $L$ is the range of pixel values. We choose the window size of 11, so the SSIM is calculated for areas of $11\times11$ pixels and then averaged over the whole image. For further details of the SSIM algorithm, see \cite{wang2004image}. The lights and shades are considered as structural features which would be included in the calculation of SSIM. Therefore, SSIM's emphasis on structural similarity could help foster the generation of realistic lights and shades. 

SSIM is defined in the range [-1, 1]. Since higher SSIM means higher similarity between the two pictures, we define SSIM loss by inverting its value:

\begin{equation}
L_{SSIM}(G)=E(1-SSIM(Y, Y^{'}))
\end{equation}

Then, we plugged in $L_{SSIM}$ into the revised model loss:

\begin{equation}
L_{revised}=\mathop{min}\limits_{G}\mathop{max}\limits_{D}L_{CGAN}(G, D)+k\cdot L_{SSIM}(G)
\end{equation}

We changed the value of the coefficient to $k=20$ in this study so that the weight of the second term in Eq. (10) remains approximately the same as the one in Eq. (5).

\subsection{Training set and validation set}

We split the 246078 pairs of data into the training set and the validation set. We noticed that randomly selecting the data pairs for the validation set would cause model leakage from the training set to the validation set. Data pairs observed 10 minutes apart would look similar to each other, so if one image is in the training set and the next image is in the validation test, the improvement of model performance on the first image might also help improve the model performance on the second image. This would cause an overestimation of model performance on the validation set. To prevent this, we retrieved the Local Mean Time (LMT) of each data pair and grouped them by the date. We then randomly selected 10\% of the dates and used all data pairs on the selected dates as the validation set. This ensures that the data pairs in the training set and the validation set are at least 6 hours apart because there is no data pair from areas with solar zenith angles above 80°. Given that satellite images at local scales change drastically within periods of 6 hours, this would avoid leakage from the training set to the validation set. We obtained a total of 23619 data pairs in the validation set and 222459 data pairs in the training set. Among the 23619 data pairs, 17757 pairs are from full-disk data, and 5862 pairs are from target area data.

This study includes additional nighttime datasets created using both VIIRS and AHI data. The nighttime datasets naturally belong to the validation set because the data pairs from the training set are all from the daytime. The VIIRS passes always occur shortly after midnight in local time, which makes them at least 5 hours from the time when daytime VIS images in the training set are available. This avoids possible model leakage from the training set to the VIIRS validation set.

\subsection{Model validation and statistical comparison}

We divided model validation into two parts: daytime validation on AHI datasets and cross-satellite nighttime validation using AHI and VIIRS datasets.

We performed validation on AHI datasets by applying our model to a total of 23619 data pairs in the validation set and making statistical comparisons between model-generated images and real VIS images. Apart from averaging the statistical results over the whole validation set, we divided the AHI validation set into different scenarios to examine whether the model is consistently robust under all conditions. First, we performed model validation on data groups with different solar zenith angles and solar azimuth angles. Second, we evaluated model performance in different geographical regions in the full-disk image.

Since our model is eventually used for simulating nighttime visible imagery, we tested the model performance during nighttime using both AHI data and VIIRS DNB data. This approach of using DNB imagery as a substitute for visible imagery at night is also adopted by \cite{pasillas2024turning} and \cite{chirokova2023proxyvis}.

\begin{figure*}[htbp]
	\centering
	\includegraphics[width=6.5in]{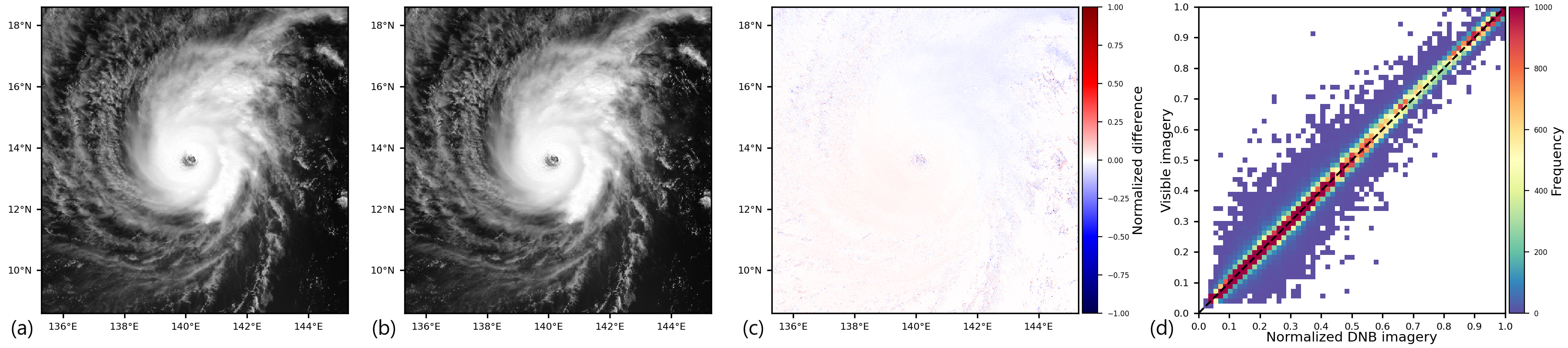}
	\caption{(a) Normalized DNB imagery, (b) I1 visible imagery, (c) difference image, and (d) 2D histogram of Typhoon Wutip (2019) at 03:42 UTC on February 25. The color scale in (c) matches Fig. 1.}
\end{figure*}

We selected the AHI full-disk data that are closest to the time of nighttime VIIRS passes, which occurs once a day for a given satellite. The time difference between the VIIRS pass and the AHI observation is within 5 minutes. The model generates images by using AHI IR data as inputs. To provide appropriate input for the ``sunlight directions", we retrieved lunar zenith angles and lunar azimuth angles of the VIIRS passes to substitute solar zenith angles and solar azimuth angles as model input. The satellite viewing angles are still retrieved from AHI data. We cropped the AHI full-disk data to 500×500 pixels and processed the DNB radiance into the same geographical range and the same resolution. Then, we normalized the radiance in 2 steps: First, we adopted the simplified High and Near Constant Contrast (HNCC) approach \cite{zinke2017simplified, martin_raspaud_2023_10400258} which neutralizes the effect of changing lunar phase; Second, the brightness is normalized by a scale factor as follows:

\begin{equation}
	x_i^{'} = x_i \cdot \frac{\mu_y}{\mu_x}
\end{equation}
where $x_i$ and $x_i^{'}$ are the brightness of individual pixels in DNB imagery before and after the scaling, and $\mu_x$ and $\mu_y$ refer to the mean values of the AI-generated imagery and the DNB imagery before the scaling.

We are aware of possible differences between the normalized DNB imagery and real visible imagery, which could be caused by the following factors:

\begin{enumerate}
	\item Difference in spectral properties of observing bands
	\item Difference between sunlight and moonlight
	\item Parallax between satellites
	\item Airglow
	\item Moon glare
	\item Lightning and artificial light
\end{enumerate}

Here we demonstrate that the impacts of 1) and 2) are minimal, and we performed quality control to minimize the difference caused by the other four factors.

VIIRS DNB has a central wavelength of 700 nm and 200 nm bandwidth, while AHI Band03 has a 645 nm central wavelength and 30 nm bandwidth. To assess the impact of differences in spectral properties, we compared VIIRS DNB with VIIRS I1, which closely matches AHI Band03 (640 nm central wavelength, 40 nm bandwidth). Fig. 6 shows a high similarity between VIIRS DNB and I1 (correlation coefficient: 0.997), indicating that the spectral differences are negligible.

Both the moon and the sun are light sources with sizes of $\sim0.5^\circ$, which can be simply treated as point sources with different emitting power. The moon also has a similar spectrum to the sun at DNB's wavelength range despite being slightly modified by the reflection spectra \cite{ohtake2013one}. The consistency of lunar illumination is also supported by \cite{min2017investigation}, which indicates minimal differences (less than 0.05\%) for cloud reflectance at different lunar phase angles.

The parallax is caused by different viewing angles between the SSO satellites carrying VIIRS and the geostationary satellites carrying AHI, which is considered a major source of difference between the AI-generated imagery and the normalized DNB imagery. The parallax would cause the same cloud to have different projected longitude and latitude for different satellites. For clouds close to the tropopause in tropical regions, a $\sim$60$^\circ$ difference in satellite viewing angle leads to a parallax displacement of $\sim$20 km. The displacement causes serious mismatches between the DNB images and the AI-generated images, negatively impacting the results of statistical comparison. To reduce the impact of parallax, we adopted VIIRS passes near the equator and matched the viewing angle of the two satellites at the center of the image. However, as the viewing angles of VIIRS vary rapidly with geographical position, minor parallax still occurs at the edge of the images. This issue is further discussed in Section IV.\textit{B}.

DNB images also detect airglow, which has a typical radiance of $2\times10^{-10} W/cm^{2}sr^{-1}$ near the nadir of DNB scan \cite{uprety2017improving}. We restricted the lunar zenith angles to below $60^\circ$ which corresponds to radiance of $\sim7\times10^{-8} W/cm^{2}sr^{-1}$ for high reflectivity clouds \cite{cao2019radiometric} on average. This ensures that the moonlight is bright enough so that the reflectivity contribution of airglow is below 3\%.

The presence of the moon glare affects model validation. Since our model uses the satellite viewing angles of AHI data, it cannot simulate the glares observed by VIIRS. We calculated the angular distance between the center of the image and where the glare would be the most intense as follows: 

\begin{equation}
	\cos\alpha = \cos\theta_1\cos\theta_2+\sin\theta_1\sin\theta_2\cos(\phi_2-\phi_1+180)
\end{equation}
where $\alpha$ is the angular distance to the glare, and $\theta_1$, $\theta_2$, $\phi_1$, and $\phi_2$ refers to solar zenith angle, satellite zenith angle, solar azimuth angle, and satellite azimuth angle respectively. To make the images free of glare, we excluded the data where the angular distance is below $45^\circ$.

The DNB imagery might also contain lightning and artificial light that do not exist in normal daytime VIS images. Thus, we perform manual quality control to exclude these DNB images.

We used the VIIRS DNB data of NOAA-20 from February 27, 2018 to November 1, 2018. After quality control, we obtained a total of 36 DNB passes with corresponding AHI observations. Fig. 7 shows a flow chart summarizing the procedures of nighttime model validation.

\begin{figure}[htbp]
	\centering
	\includegraphics[width=3.5in]{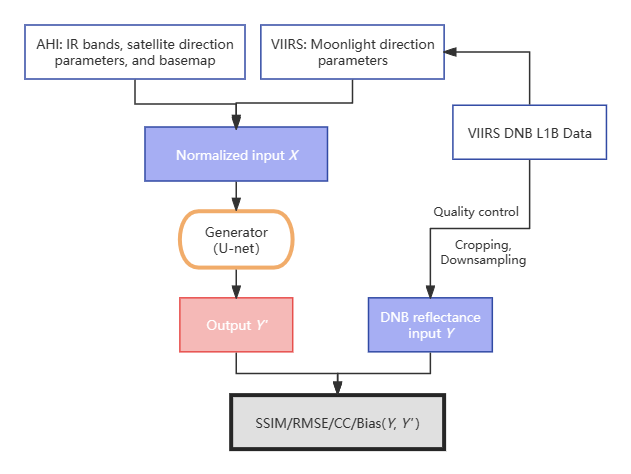}
	\caption{Flow chart of nighttime model validation using AHI and VIIRS data.}
\end{figure} 

We adopted the following statistical comparison algorithms: SSIM (see Eq. (8)), Peak Signal-to-Noise Ratio (PSNR), Root Mean Square Error (RMSE), Correlation Coefficient (CC), and Bias:

\begin{equation}
PSNR(x, y)=10\cdot\log_{10}\frac{L^2}{\frac{1}{N}\sum_{i}^{N}(x_i-y_i)^2}
\end{equation}

\begin{equation}
RMSE(x, y)=\sqrt{\frac{1}{N}\sum_{i}^{N}(x_i-y_i)^2}
\end{equation}

\begin{equation}
CC(x, y)=\frac{\sum_{i}^{N}(x_i-\mu_x)(y_i-\mu_y)}{\sqrt{\sum_{i}^{N}(x_i-\mu_x)^2}\sqrt{\sum_{i}^{N}(y_i-\mu_y)^2}}
\end{equation}

\begin{equation}
Bias(x, y)=\frac{1}{N}\sum_{i}^{N}(x_i-y_i)
\end{equation}
where $x$ and $y$ refer to the two input images, $x_i$ and $y_i$ refer to the values of individual pixels, $\mu_x$ and $\mu_y$ refer to the mean value of pixels in $x$ and $y$, $L$ represents the maximum possible pixel value of the image, and $N$ represents the number of pixels.

\section{Results}

The training process lasted 600 epochs, and each epoch contains 9270 iterations. The model performance on the validation set leveled off after epoch 500, and we adopted the result at epoch 578 (around 5360000 iterations) to be our final model for validation and application.

\begin{figure}[htbp]
	\centering
	\includegraphics[width=3.5in]{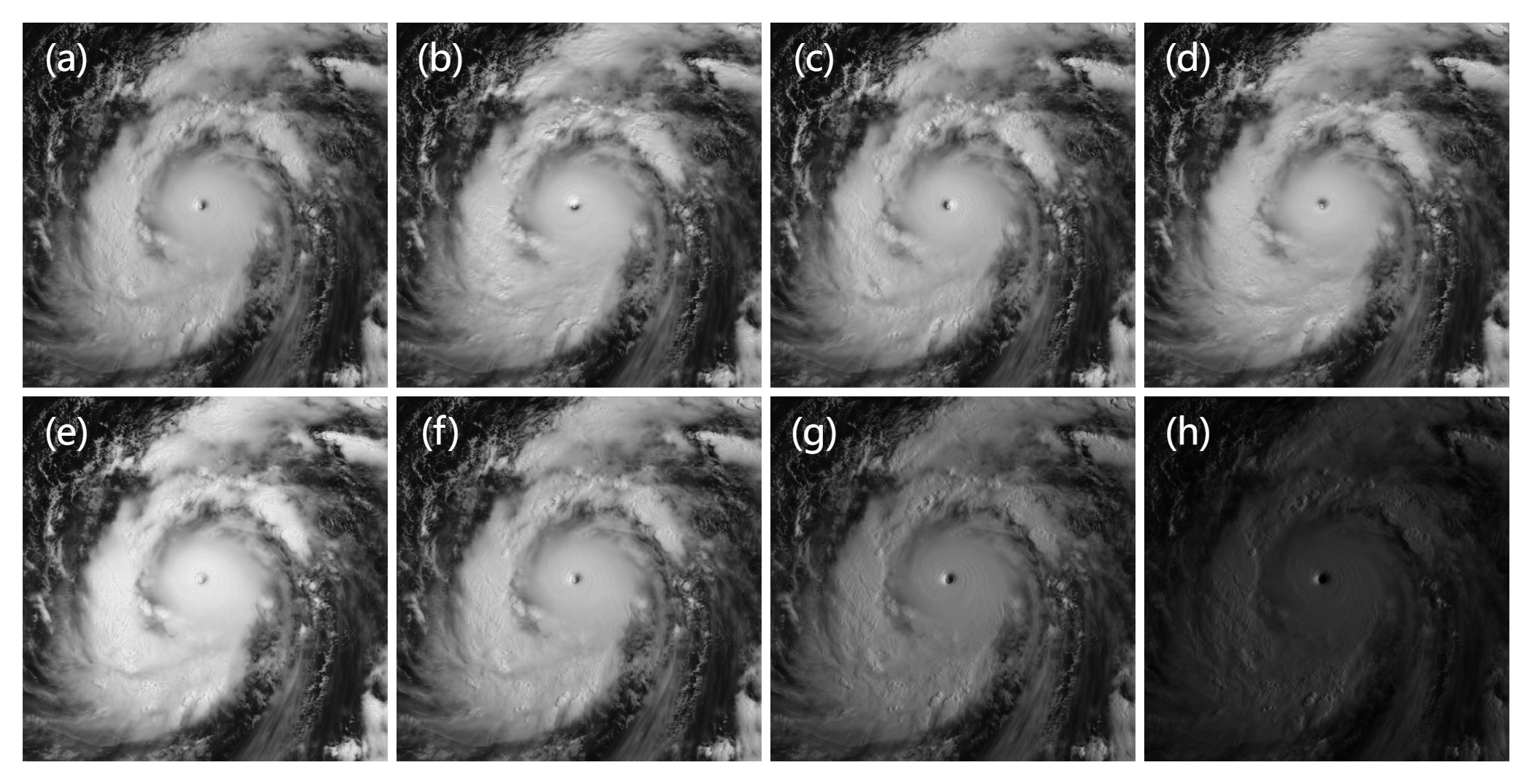}
	\caption{AI-generated images of Typhoon Bolaven (2023) at 11:30 UTC (nighttime) on October 11 with different sunlight direction parameters. Image (a), (b), (c), and (d) have a fixed solar zenith angle of 35° and solar azimuth angles of 90°, 180°, -90°, and 0° respectively. Image (e), (f), (g), and (h) have a fixed solar azimuth angle of 90° and solar zenith angles of 15°, 35°, 55°, and 75° respectively.}
\end{figure}

An outstanding feature of our model is that the virtual sunlight direction of the AI-generated image could be arbitrarily manipulated by changing the input of the solar zenith angle and the solar azimuth angle. This means that we could use the model to simulate VIS images as if they were at any time of the day. Fig. 8 presents a series of images generated from the same IR input but with different solar zenith angles and solar azimuth angles. Our model provides realistic simulations of clouds under changing sunlight directions.

\begin{figure*}[htbp]
	\centering
	\includegraphics[width=6in]{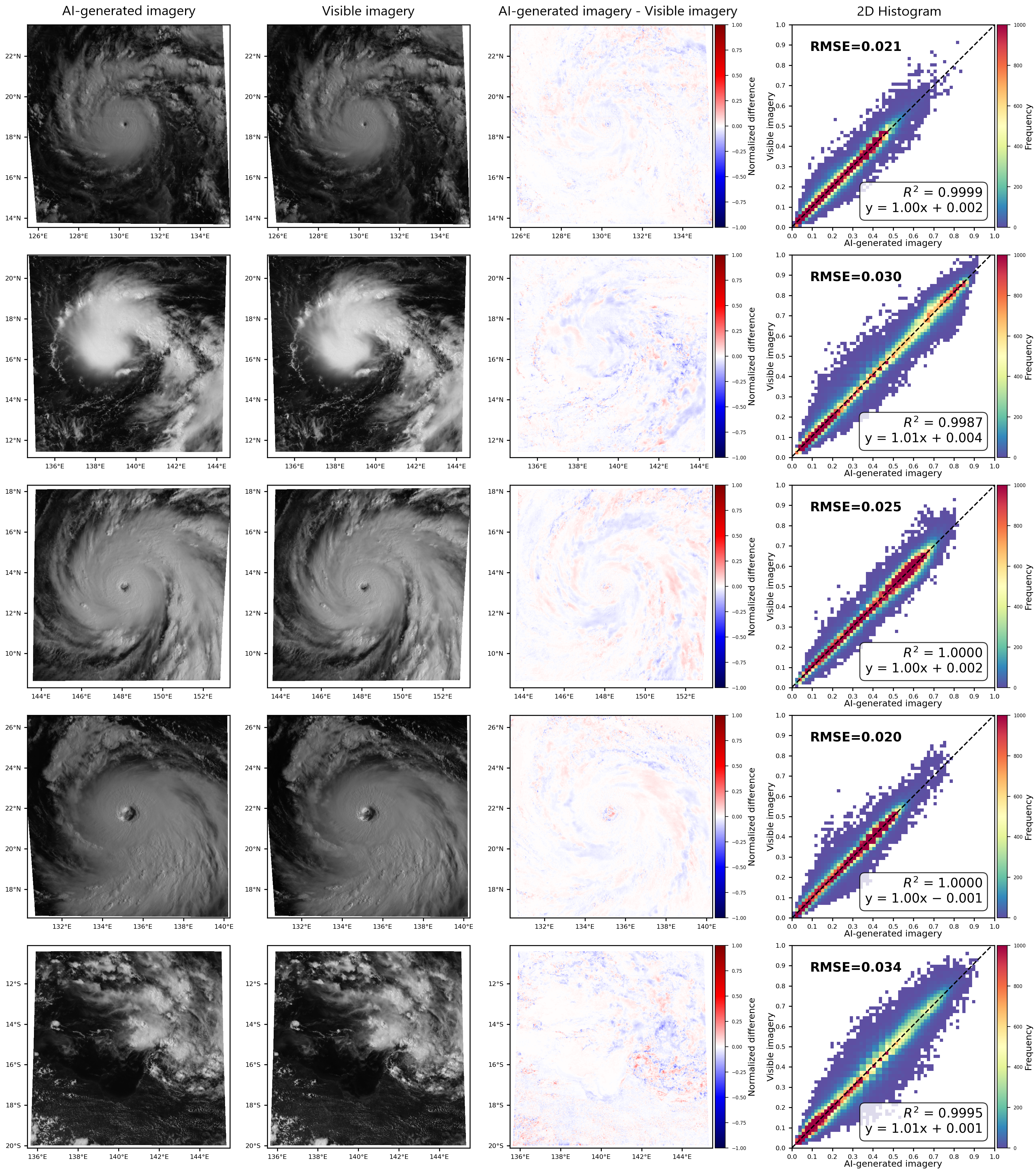}
	\caption{Sample daytime image pairs and their corresponding difference images and 2D histograms. The color scale in the third column matches Fig. 1, and the dashed lines in the fourth column are the best-fit lines of linear regression.}
\end{figure*}

\begin{table*}[htbp]
	\begin{center}
		\footnotesize
		\caption{Daytime statistical results of our model and other existing models}
		\renewcommand{\arraystretch}{1.5}
		\begin{tabular*}{0.663\linewidth}{|c|c|p{0.7cm}p{0.7cm}p{0.9cm}p{0.7cm}p{1.3cm}|c|}
			\hline
			Study & Network & SSIM & PSNR & RMSE & CC & Bias & Details\\
			\hline
			Our model & CGAN & \textbf{0.923} & \textbf{31.4} & \textbf{0.0299} & \textbf{0.991} & \textbf{0.0003} & -\\
			\hline
			\multirow{2}*{Kim \textit{et al.}\cite{kim2019nighttime}} & \multirow{2}*{CGAN} & - & - & 0.129\footnotemark[1] & 0.89 & -0.0008 & Winter\\
			& & - & - & 0.145\footnotemark[1] & 0.88 & -0.0095 & Summer\\
			\hline
			Kim \textit{et al.}\cite{kim2020impact} & CGAN & - & - & 0.105\footnotemark[1] & 0.952 & -0.0069 & - \\
			\hline
			\multirow{3}*{Harder \textit{et al.}\cite{harder2020nightvision}} & CGAN & 0.77 & - & 0.11 & - & - & -\\
			& U-Net\footnotemark[2] & 0.85 & - & 0.09 & - & - & -\\
			& U-Net$++$\footnotemark[2] & 0.86 & - & 0.07 & - & - & -\\
			\hline
			\multirow{3}*{Han \textit{et al.}\cite{han2022hypothetical}} & \multirow{3}*{CGAN} & - & - & 0.061 & 0.917 & -0.010 & Red\\
			& & - & - & 0.050 & 0.939 & -0.007 & Green\\
			& & - & - & 0.047 & 0.941 & -0.006 & Blue\\
			\hline
			Cheng \textit{et al.}\cite{cheng2022creating} & CGAN & 0.480 & 25.5\footnotemark[3] & 0.082 & - & - & -\\
			\hline
			Yan \textit{et al.}\cite{yan2023simulation} & DNN & - & - & 8.38\footnotemark[4] & - & - & -\\
			\hline
		\end{tabular*}
	\end{center}
\end{table*}

\subsection{Daytime validation performance}

The daytime validation performance is evaluated by applying 5 different statistical comparison algorithms (SSIM, PSNR, RMSE, CC, Bias) to the 23619 data pairs in the AHI validation set. Table III summarizes the values given by these algorithms averaged over 23619 data pairs and compares the result with previous studies. The data range of the images is [0, 1]. Note that different studies have chosen data that cover different areas of interest as their datasets, so the numerical values presented might not strictly represent model performance.

Among all models, our model has the best performance for all 5 statistical comparison algorithms by a large margin. The outstanding SSIM values demonstrate that our model successfully reconstructs structural details in VIS images, while the PSNR, RMSE, and CC values also suggest minimal differences between our AI-generated images and the VIS images.

Fig. 9 shows 5 AI-generated images in the validation set, their paired real VIS images, the difference images between AI-generated and VIS images, and the corresponding 2D histograms. The 5 sample images are Typhoon Nepartak (2016) at 23:10 UTC on July 5, Typhoon Trami (2018) at 04:20 UTC on September 22, Typhoon Yutu (2018) at 23:00 UTC on October 23, Typhoon Haishen (2020) at 23:00 UTC on September 3, and Cyclone Lucas (2021) at 07:40 UTC on February 6 in sequence. This covers a variety of different solar zenith angles, solar azimuth angles, seasons, latitudes, and TC intensities. The images of Cyclone Lucas demonstrate that our model is capable of simulating land regions with certain reflectivity. We performed linear regression for each of the image pairs and overlaid the best-fit lines on the 2D histograms. In all 5 cases, the best-fit lines lie very close to the $y=x$ line, with $R^2$ values exceeding 0.9995, indicating an outstanding and consistent model performance. We also noticed that the 2D histograms show very few pixels with AI-generated reflectance above 0.95, while the real VIS images do have reflectance $\sim$1 at some pixels. This might have caused the slight negative Bias at low solar zenith angles as shown in Table IV.

To evaluate the consistency of our model under different sunlight directions, we calculated the statistical results for different intervals of solar zenith angle and solar azimuth angle of the central pixel of data. The results are summarized in Table IV and Table V.

\begin{table}[h]
	\begin{center}
	\footnotesize
	\caption{Statistical results for different solar zenith angles}
	\begin{tabular}{m{2.1cm}<{\centering}m{0.8cm}<{\centering}m{0.5cm}<{\centering}m{0.5cm}<{\centering}m{0.6cm}<{\centering}m{0.4cm}<{\centering}m{0.9cm}<{\centering}}
		\toprule
		Solar zenith angle & Sample size & SSIM & PSNR & RMSE & CC & Bias\\
		\midrule
		$0^\circ\sim10^\circ$ & 582 & 0.898 & 27.5 & 0.0442 & 0.986 & -0.0052\\
		$10^\circ\sim20^\circ$ & 1830 & 0.903 & 27.8 & 0.0426 & 0.987 & -0.0047\\
		$20^\circ\sim30^\circ$ & 2614 & 0.908 & 28.4 & 0.0399 & 0.987 & -0.0021\\
		$30^\circ\sim40^\circ$ & 3347 & 0.911 & 29.2 & 0.0369 & 0.988 & -0.0003\\
		$40^\circ\sim50^\circ$ & 3688 & 0.918 & 30.3 & 0.0323 & 0.990 & 0.0009\\
		$50^\circ\sim60^\circ$ & 3839 & 0.927 & 31.8 & 0.0272 & 0.992 & 0.0021\\
		$60^\circ\sim70^\circ$ & 3955 & 0.935 & 33.6 & 0.0222 & 0.994 & 0.0024\\
		$70^\circ\sim80^\circ$ & 3764 & 0.944 & 36.0 & 0.0170 & 0.997 & 0.0012\\
		\midrule
		Total & 23619 & 0.923 & 31.4 & 0.0299 & 0.991 & 0.0003\\
		\bottomrule
	\end{tabular}
\end{center}
\end{table}

\begin{table}[h]
\begin{center}
	\footnotesize
	\caption{Statistical results for different solar azimuth angles}
	\begin{tabular}{m{2.1cm}<{\centering}m{0.8cm}<{\centering}m{0.5cm}<{\centering}m{0.5cm}<{\centering}m{0.6cm}<{\centering}m{0.4cm}<{\centering}m{0.9cm}<{\centering}}
		\toprule
		Solar azimuth angle & Sample size & SSIM & PSNR & RMSE & CC & Bias\\
		\midrule
		$0^\circ\sim40^\circ$ & 1264 & 0.919 & 30.8 & 0.0319 & 0.990 & -0.0001\\
		$40^\circ\sim80^\circ$ & 3358 & 0.929 & 32.4 & 0.0268 & 0.992 & 0.0011\\
		$80^\circ\sim120^\circ$ & 4910 & 0.924 & 31.8 & 0.0285 & 0.992 & 0.0010\\
		$120^\circ\sim160^\circ$ & 2115 & 0.911 & 29.9 & 0.0348 & 0.989 & 0.0002\\
		$160^\circ\sim-160^\circ$ & 1010 & 0.906 & 29.2 & 0.0376 & 0.987 & -0.0004\\
		$-160^\circ\sim-120^\circ$ & 2068 & 0.914 & 29.9 & 0.0347 & 0.989 & -0.0008\\
		$-120^\circ\sim-80^\circ$ & 4590 & 0.927 & 31.9 & 0.0282 & 0.992 & -0.0000\\
		$-80^\circ\sim-40^\circ$ & 3001 & 0.930 & 32.2 & 0.0272 & 0.992 & 0.0000\\
		$-40^\circ\sim0^\circ$ & 1303 & 0.918 & 30.6 & 0.0326 & 0.990 & -0.0003\\
		\midrule
		Total & 23619 & 0.923 & 31.4 & 0.0299 & 0.991 & 0.0003\\
		\bottomrule
	\end{tabular}
\end{center}
\end{table}

The statistical results show relatively consistent performance across different solar zenith angles and solar azimuth angles. The results show $SSIM\geq0.898$, $PSNR\geq27.5$, $RMSE\leq0.0442$, $CC\geq0.986$, and $-0.0052\leq Bias\leq0.0024$ for all intervals of angles. We noticed that the model performance is slightly better at high solar zenith angles and slightly worse at low solar zenith angles. This could be attributed to that the darker high solar zenith angle images contain fewer features, making them easier to simulate; besides, our datasets contain fewer data pairs at low zenith angles, which could weaken the model performance at these angles; the occurrence of glares caused by the direct reflection of sunlight (also discussed in \cite{yan2023simulation}) on the sea might also negatively impact the statistics at low zenith angles. There is a similar variation of model performance in Table V, showing that the results are slightly better around $\theta=\pm90^\circ$ and slightly worse around $\theta=\pm180^\circ$ and $\theta=0^\circ$. This could be attributed to the lack of data pairs around $\theta=\pm180^\circ$ and $\theta=0^\circ$ and the fact that data pairs with such angles usually have low solar azimuth angles.

The statistical results indicate our model's universal application over different sunlight direction parameters. Considering the statistical results and the available information in the images, we would recommend adopting solar zenith angles between $30^\circ$ and $60^\circ$ and solar azimuth angles around $\pm80^\circ$ for operational monitoring of TCs.

We also evaluated the model performance in different geographical regions in the full-disk image. Fig. 10 shows the full-disk image stacked by 121 blocks of AI-generated images and the full-disk VIS image at 02:30 UTC on October 11, 2023. In this example, the full-disk image stacked by AI-generated images accurately simulates the VIS image, and the gaps between neighboring blocks are barely distinguishable.

\begin{figure}[htbp]
	\centering
	\includegraphics[width=3.5in]{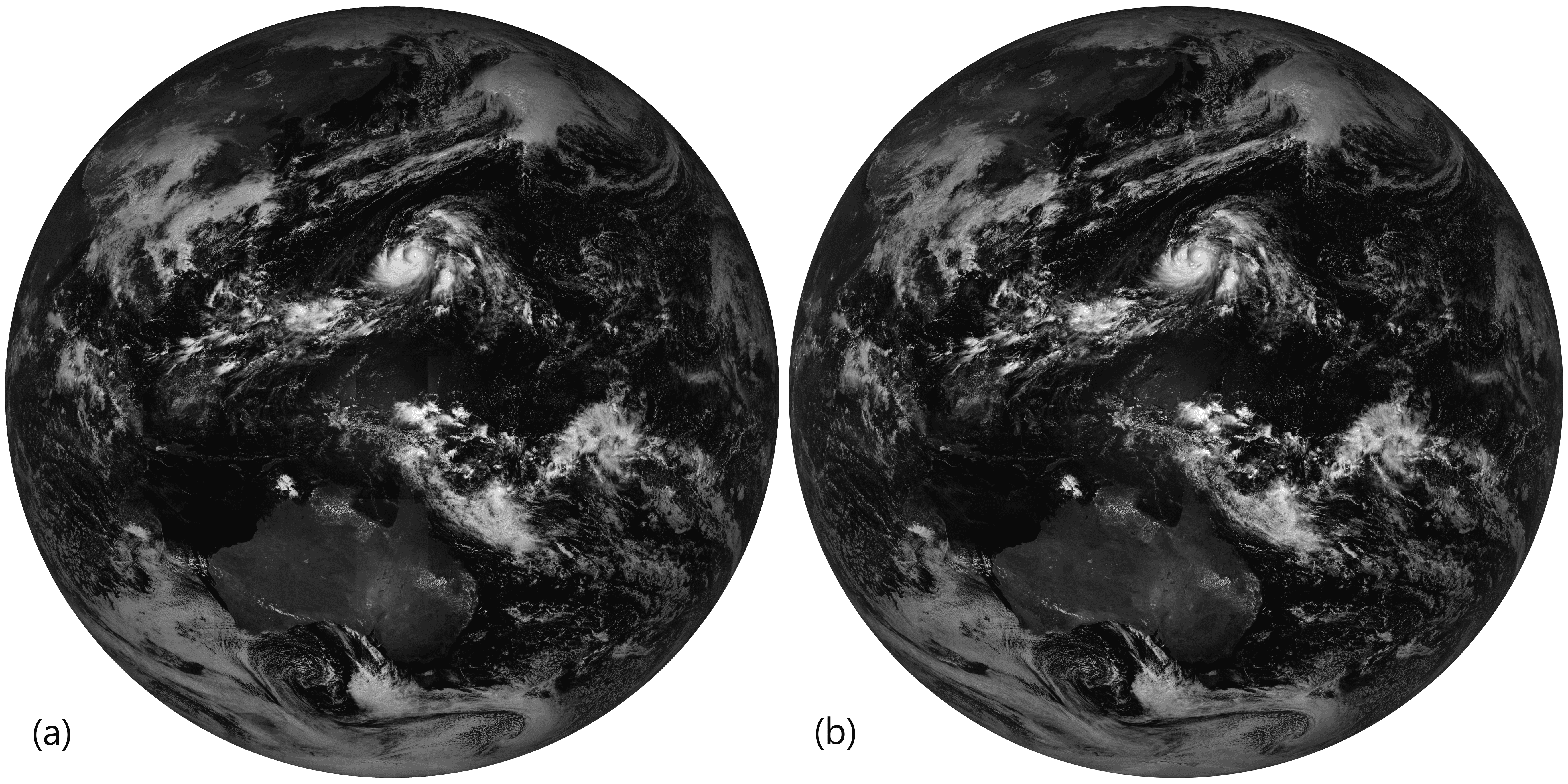}
	\caption{(a) AI-generated image stacked by the blocks and (b) full-disk VIS image at 02:30 UTC on October 11, 2023.}
\end{figure}

\begin{figure}[htbp]
	\centering
	\includegraphics[width=3.5in]{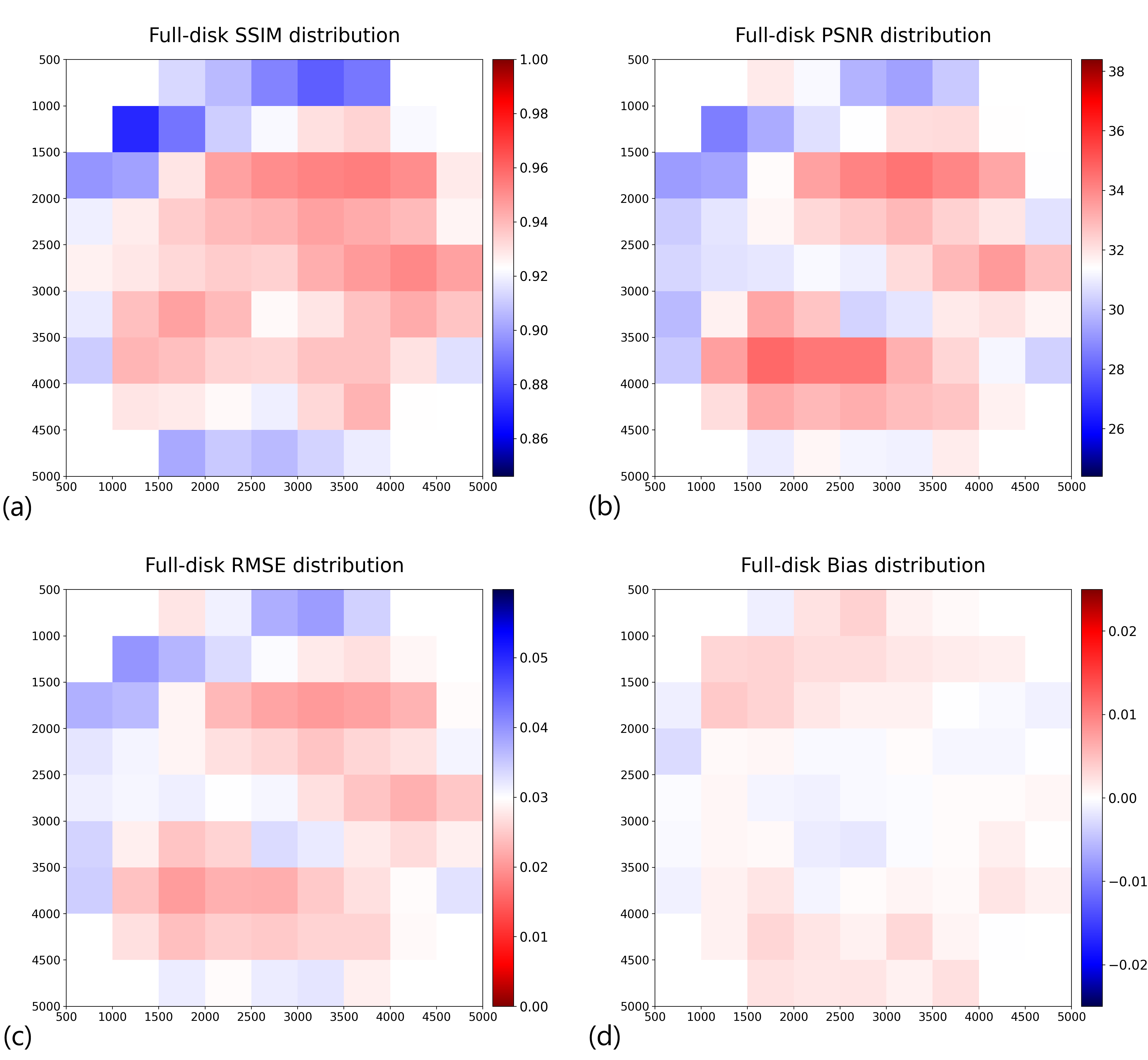}
	\caption{Distribution of averaged (a) SSIM, (b) PSNR, (c) RMSE, and (d) Bias of each block. The x-axis and y-axis represent the column and line of full-disk data.}
\end{figure}

We divided 15,716 data pairs cropped from the full-disk data in the validation set into 69 blocks and calculated the average statistical results for each block. Fig. 11 shows the distribution of averaged SSIM, RMSE, and Bias. In Fig. 11a, 11b, and 11c, the red color represents the blocks with better performance than the average, while the blue color indicates worse performance. The results show $SSIM\geq0.871$, $PSNR\geq28.6$, $RMSE\leq0.0397$, $CC\geq0.982$, and $-0.0028\leq Bias\leq0.0043$ for all blocks. The model performance is generally consistent across the full-disk observation despite being slightly better for open-sea areas. The satellite viewing angles in the input dataset contribute to the consistency of the model performance across the full-disk image, as it improves the simulations when the data is near the edge or affected by the glare near the equator. The results in this section demonstrate that our model could be readily applied to both target area observations centering at TCs and full-disk observations.

\footnotetext[1]{The RMSE and Bias values in \cite{kim2019nighttime} and \cite{kim2020impact} are normalized from [0, 255] to [0, 1].}
\footnotetext[2]{The U-Net and U-net++ models have relatively inconsistent performance at night according to \cite{harder2020nightvision}.}
\footnotetext[3]{We converted the PSNR value in \cite{cheng2022creating} into the unit of dB.}
\footnotetext[4]{The data range used to calculate the RMSE is not clearly indicated in \cite{yan2023simulation}.}

\subsection{Nighttime validation performance}

The nighttime validation performance is evaluated by applying the statistical comparison algorithms to the image pairs of VIIRS DNB normalized radiance and their corresponding AI-generated reflectance derived from AHI data. The scaling process introduced in Section III.\textit{C} reduces the Bias between the two images to 0. Thus, only four statistical comparison algorithms (SSIM, PSNR, RMSE, CC) are presented in this section. The inability to evaluate Bias for the model at night is further discussed in Section V.

The statistical results yielded from the comparison between 36 pairs of normalized DNB radiance and AI-generated reflectance are summarized in Table VI and compared to the result in \cite{pasillas2024turning}, which is currently the only other study that has performed a quantitative nighttime validation. Although SSIM and PSNR were not evaluated in \cite{pasillas2024turning}, the RMSE and CC values of our model significantly outperform those of existing models.

\begin{table}[htbp]
	\begin{center}
	\footnotesize
	\caption{Nighttime statistical results of our model and the existing model}
	\renewcommand{\arraystretch}{1.5}
	\begin{tabular}{cccccc}
		\toprule
		Study & Network & SSIM\footnotemark[5] & PSNR & RMSE & CC\\
		\midrule
	    Our model & CGAN & 0.860 & 25.4 & 0.0551 & 0.977\\
	    
	    Pasillas \textit{et al.}\cite{pasillas2024turning} & FNN & - & - & 0.0933\footnotemark[6] & 0.91\\
		\bottomrule
	\end{tabular}
\end{center}
\end{table}
\footnotetext[5]{To reduce the impact of parallax on SSIM values, we increased the window size by about 10 times from 11 to 111 pixels.}
\footnotetext[6]{We adopted the best statistical result among all combinations of models and latitude ranges presented in \cite{pasillas2024turning}.}

Fig. 12 presents an image pair of AI-generated imagery and Normalized DNB imagery of clouds near the equator. A variety of clouds including cumulonimbus, cumulus, and cirrus detected by DNB imagery are accurately reproduced in the AI-generated imagery, with precise simulation of light and shadow based on the lunar zenith and azimuth angles.

Fig. 14 presents AI-generated imagery and normalized DNB imagery in pairs, their difference images, and the 2D histograms. To minimize the parallax between satellites, images near the equator were chosen. Since most TCs are away from the equator, no TCs were included. The sample images are at 15:18 UTC on February 28, 2018, 15:10 UTC on April 28, 2018, 14:29 UTC on May 27, 2018, 14:36 UTC on August 25, and 14:12 UTC on September 22, 2018 in sequence. The image pairs have a RMSE varying from 0.036 to 0.071 depending on the complexity of cloud features. The linear regressions yield $R^2$ values above 0.9944 for all 5 image pairs, showing a strong correspondence between AI-generated and normalized DNB images. The best-fit lines have slopes slightly below 1, likely because the residual moon glare slightly increases the measured reflectance of the clear sea surface.

\begin{figure}[htbp]
	\centering
	\includegraphics[width=3.5in]{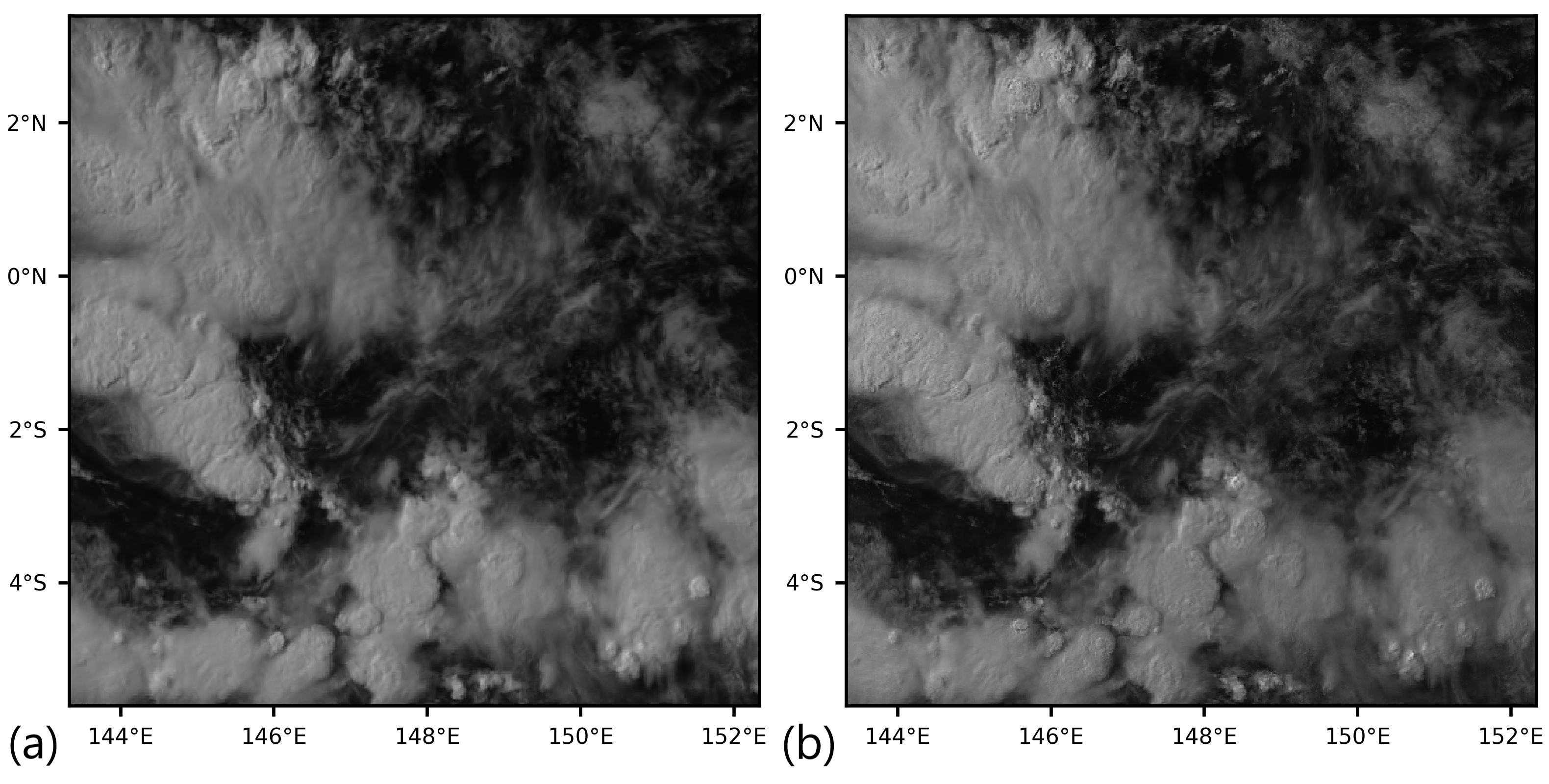}
	\caption{(a) AI-generated imagery and (b) Normalized DNB imagery of clouds near the equator at 15:35 UTC on February 27, 2018.}
\end{figure}

\begin{figure}[htbp]
	\centering
	\includegraphics[width=3.2in]{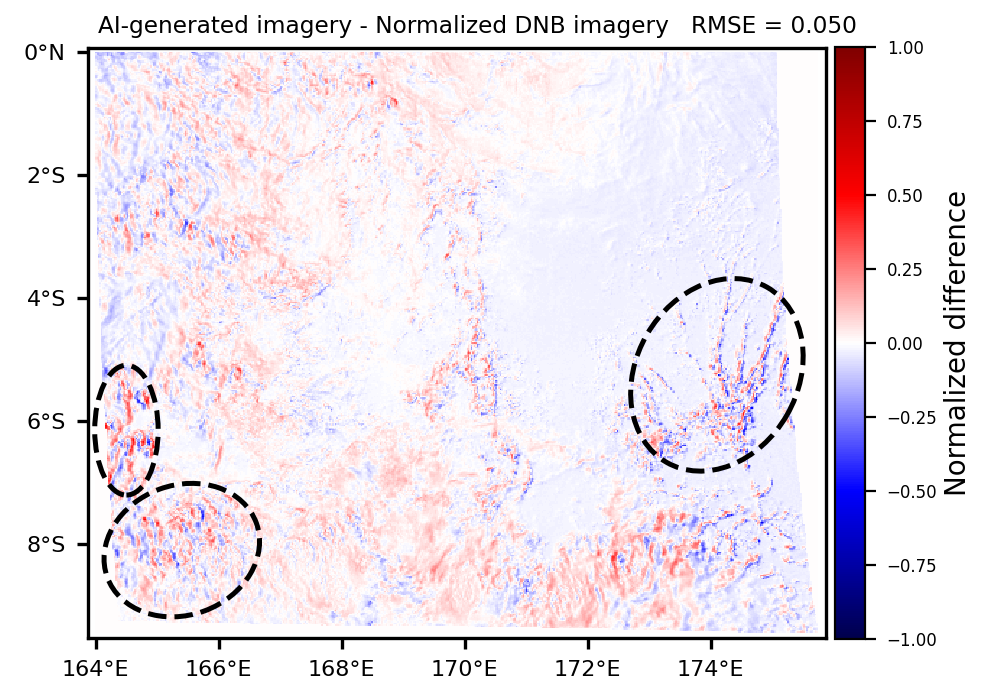}
	\caption{The sample difference image at 14:12 UTC on September 22, 2018, with three regions affected by parallax circled.}
\end{figure}

\begin{figure*}[htbp]
	\centering
	\includegraphics[width=6in]{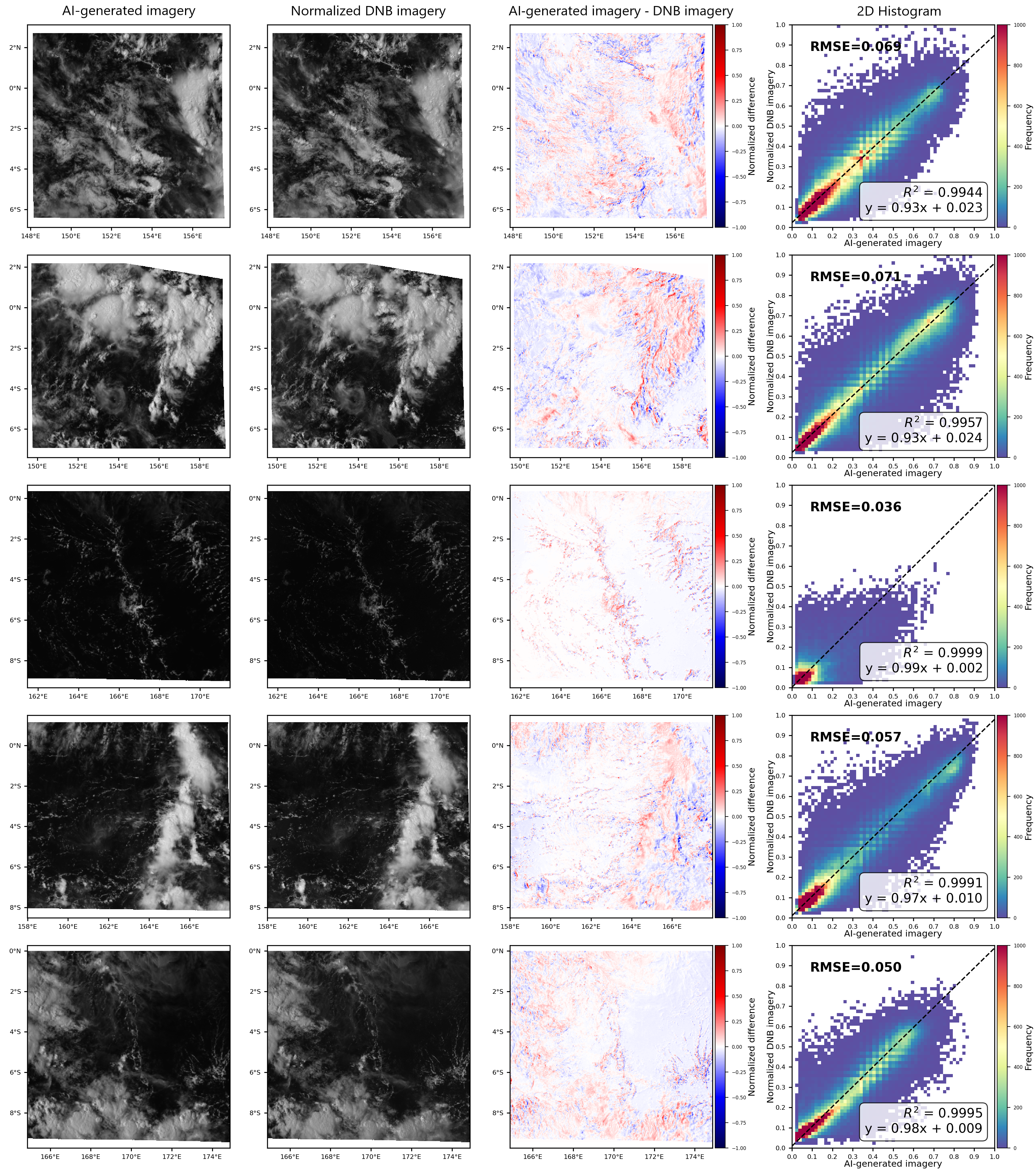}
	\caption{Sample nighttime image pairs and their corresponding difference images and 2D histograms. The color scale in the third column matches Fig. 1, and the dashed lines in the fourth column are the best-fit lines of linear regression.}
\end{figure*}

The nighttime statistical result in Table VI is slightly worse than the daytime statistical result in Table III. Besides the minor effect of moon glare, we primarily attribute this decline to parallax at the image edges, which is caused by VIIRS’s rapidly varying viewing angles. Fig. 13 zooms in on the difference image from the fifth row of Fig. 14, in which several regions affected by the parallax are identified. The mismatch of images caused by parallax creates streak-like patterns on the difference image, contributing a considerable part of the RMSE between the two images. While the parallax negatively impacts the statistical results, the cloud features remain accurately simulated, albeit in slightly different locations. Additionally, although the geographical location of clouds simulated using AHI IR data mismatches that of VIIRS data, it is still consistent with AHI VIS data. Therefore, the existence of parallax in cross-satellite nighttime validation does not affect the accuracy and utility of AI-generated images.

In conclusion, despite the statistical results being slightly worse at night, the AI-generated images remain highly accurate, demonstrating successful model generalization from daytime to nighttime. We attribute this success to the relatively low diurnal temperature variation in tropical ocean regions, which supports the fundamental assumption that the IR bands have consistent $T_b$ from daytime and nighttime.

\subsection{Applications}

This section illustrates the application of our model in TC monitoring under two different scenarios: exposed Low-Level Circulation Center (LLCC) and early stages of eye formation.

Relatively weak TCs under vertical wind shear (VWS) might not have any convection directly covering their LLCC. The center of such TCs could be determined by the curvature of the low-level clouds surrounding the LLCC using VIS images. However, low-level clouds are usually difficult to identify in IR images as shown in Fig. 15a, which makes positioning harder at night. In Fig. 15b, the AI-generated image simulates the low-level clouds surrounding the LLCC, which is indicated by the orange cross. The positioning using the AI-generated image is verified by the ASCAT Coastal Wind data, which shows that the direction of wind follows the curvature of the clouds simulated by our model, and the center of TC corresponds to the center of curvature of the low-level clouds.

\begin{figure*}[htbp]
	\centering
	\includegraphics[width=6in]{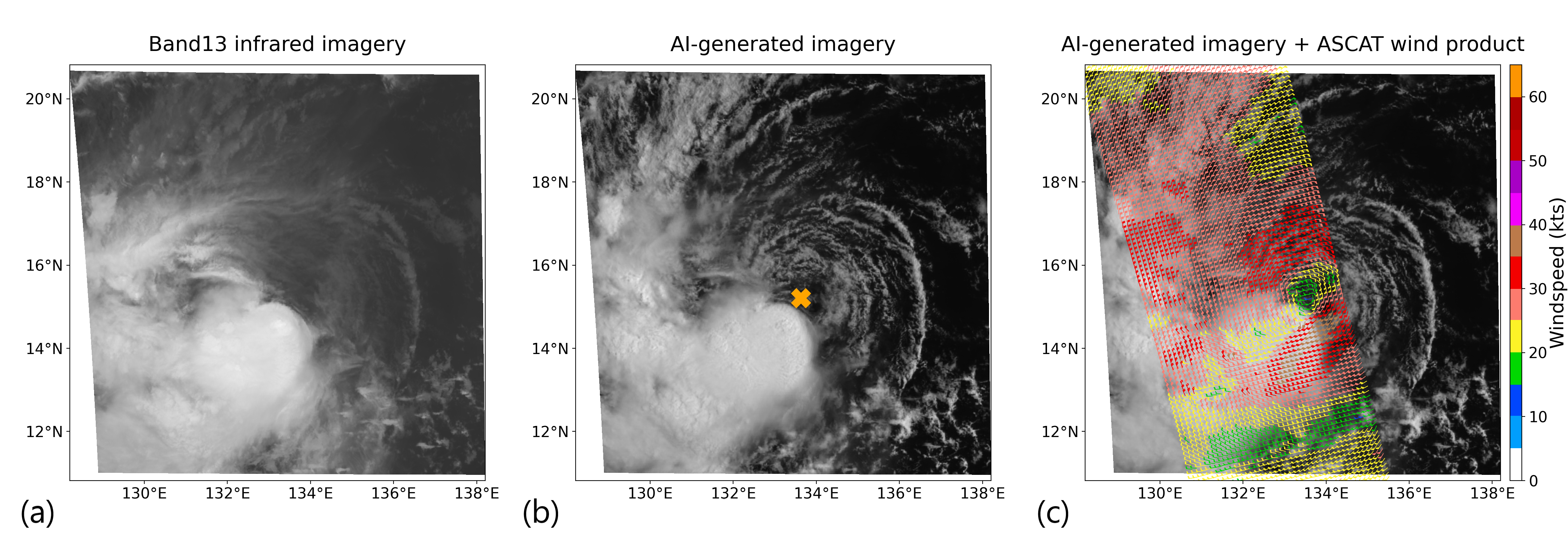}
	\caption{(a) AHI Band13 IR imagery, (b) AI-generated imagery, and (c) AI-generated imagery overlapped by ASCAT wind product of Typhoon Koppu (2015) at 12:10 UTC on October 14. The orange cross indicates the position of the LLCC.}
\end{figure*}

\begin{figure*}[htbp]
	\centering
	\includegraphics[width=6in]{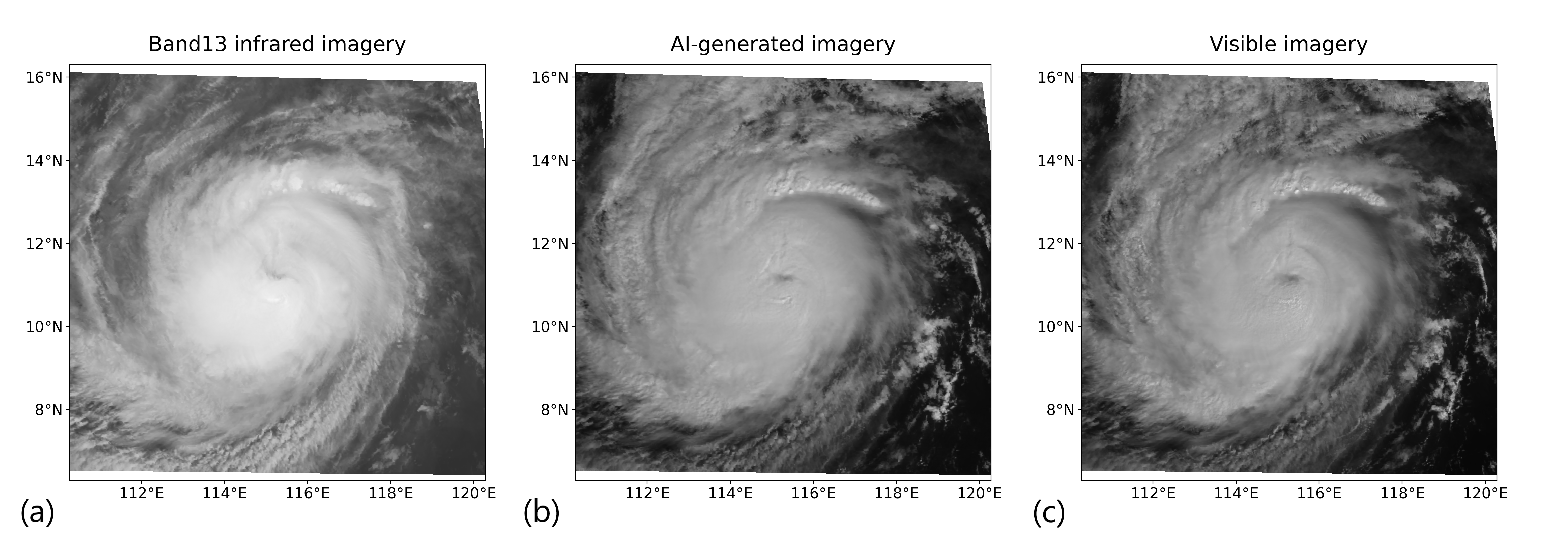}
	\caption{(a) Band13 IR imagery, (b) AI-generated imagery, and (c) visible imagery of Typhoon Rai (2021) at 02:20 UTC on December 18. The blurry low-reflectivity region at the center indicates the eye.}
\end{figure*}

While the ASCAT can only provide one pass for a particular typhoon each night, our model generates continuous images at time intervals of 2.5 minutes (target area) or 10 minutes (full-disk). Therefore, our model could enhance the continuous positioning of weak TCs for operational purposes.

Eye formation is a crucial process in TC intensification. Two types of eye formation are proposed by \cite{chen2022two}: clearing formation (CF) and banding formation (BF). CF is usually related to the Rapid Intensification (RI) of TCs. Thick cirrus remain in the eye during the early stages of CF, which makes it relatively difficult to identify the eye in IR images. However, with a much lower imaginary index of refraction for ice particles \cite{warren2008optical}, visible light could penetrate the cirrus in the forming eye. Therefore, it is easier to identify the eye during CF in VIS images than in IR images, and the identification of forming eyes significantly contributes to TC intensity forecasts.

Fig. 16 shows the IR image, the AI-generated image, and the VIS image of Typhoon Rai (2021) at the early stages of eye formation. While the eye is nearly unidentifiable in the IR image, we could see a blurry region of low reflectivity at the center of both the AI-generated image and the VIS image, which is the eye of Typhoon Rai filled with relatively thick cirrus. The images are from the AHI validation set, so this is a valid representation of model performance. Therefore, our model could be applied to nighttime and help the identification of forming eyes when VIS images are unavailable.

\section{Discussion}

This study presents a CGAN model that converts IR data, direction parameters of the sun and the satellite, and land basemap into highly accurate AI-generated VIS reflectance. As a result, the CGAN model significantly enhances nighttime TC monitoring when VIS images are unavailable. Currently, we provide real-time AI-generated images of TCs to the public (https://ai-vis.dapiya.top). The complete source code and datasets used in this study are publicly available at our GitHub repository (https://github.com/Dapiya/AI-VIS).

Among all established DL approaches to simulate nighttime visible imagery, our model achieves the best statistical results for daytime model validation. We yielded the highest SSIM of 0.923, the highest PSNR of 31.4, the lowest RMSE of 0.0299, the highest CC of 0.991, and the smallest Bias of 0.0003. The excellent model performance can be attributed to various aspects of our model design. First, we adopted the SSIM loss instead of the L1 loss in the CGAN model which encourages the simulation of high-frequency information such as lights and shades on the cloud top; Second, we selected appropriate IR bands by examining their spectral characteristics instead of using their correlation coefficient with VIS reflectance; Third, we adopted 4 direction parameters (solar zenith angle, solar azimuth angle, satellite zenith angle, and satellite azimuth angle) and a land basemap as extra channels in the input datasets. The solar zenith angle and the solar azimuth angle contribute to the accurate simulation of lights and shades of the clouds. Additionally, instead of using images at the same time each day, the solar zenith angle and the solar azimuth angle enable us to include far more image pairs in the training set, significantly improving the generalization from the training set to the validation set. The satellite zenith angle, the satellite azimuth angle, and the land basemap also provide crucial information for the simulation of VIS images, enabling better simulations for the edge of full-disk observation, glare-affected areas, and land areas.

The cross-satellite nighttime model validation using VIIRS DNB data demonstrates the successful generalization of our model from daytime to nighttime. Despite being slightly affected by the parallax, the statistical results are outstanding compared to other existing models. Therefore, our model can be readily applied to nighttime IR observations and provide real-time nighttime monitoring of the earth.

Beyond statistical accuracy, our model offers several additional advantages in the practical monitoring of weather systems. First, our model is capable of simulating VIS images with arbitrary virtual sunlight directions. VIS images with different solar zenith angles have various useful properties. For example, images with low solar zenith angles show penetrate optically thin clouds better, helping visualize cirrus-covered cloud features; images with high solar zenith angles show higher contrast of lights and shades, facilitating the identification of developing thunderstorms. As a result, a combination of multiple AI-generated images assuming various sunlight direction would further enhance the effectiveness of nighttime weather monitoring.

In addition, our model has the highest resolution ($\sim$2 km) among all current models. With the relatively small size ($\sim$1000 km) of data pairs, we constructed the model at the original resolution of IR bands. We are also currently working on a Super-Resolution model based on Real-ESRGAN \cite{wang2021real} which further increases the resolution of AI-generated images from 2 km to 0.5 km.

While most contemporary DL approaches more or less face the challenge of computational demand, our data pair size also enables the model to be applied operationally with minimal computational costs. Running on a personal computer with NVIDIA GeForce RTX 4060 Laptop GPU and AMD Ryzen 7 7840H CPU, the output image can be generated from target area data within 0.5 seconds, and the model inference only takes $\sim$0.05 seconds.

Our model can be readily applied to various observation modes of different geostationary satellites. Full-disk AI-generated reflectance can be obtained by stacking individual blocks of cropped images as shown in Fig. 10. Since the imagers on GK-2A and GOES-16/17/18 have similar spectral characteristics to AHI on Himawari-8/9, our model can also be applied on these satellites with a different land basemap input, enabling global coverage of AI-generated reflectance. 

We identified some limitations of our current work presented in this study. While our model shows robust performance across various cloud scenarios, its accuracy is slightly worse in regions covered by thick cirrus. Since all adopted IR bands cannot effectively penetrate thick cirrus, there is limited physical information available from the regions below. This issue can be addressed in future studies by adopting other imagers' IR bands with a lower imaginary index of refraction.

Our nighttime model validation suffers from the parallax between satellites and the inability to evaluate the Bias. We plan to evaluate our model with feature-based metrics such as Scale-Invariant Feature Transform (SIFT) \cite{lowe2004distinctive} that tolerate minor parallax between images, which enable us to compare the model performance between daytime and nighttime more effectively. The current nighttime model validation also fails to calculate the Bias of the model because of the scaling process to generate normalized DNB imagery. This issue could be addressed in the future by adopting the algorithm proposed in \cite{liang2014improved}, which directly calculates pseudo-albedo from VIIRS DNB radiance.

Although our model has already achieved impressive statistical results, it is rather challenging to physically interpret the contribution of each input channel. The study adopting the traditional multiple linear regressions yielded the contribution of each input channel directly \cite{chirokova2023proxyvis}, and the study adopting a DNN model presented the weights of each channel in the full link network layer \cite{yan2023simulation}. To the best of our knowledge, however, none of the studies adopting the CGAN model has interpreted the contribution of each channel, as the physical interpretation of GANs is naturally more challenging in comparison. To address the lack of physical interpretations of what the model has learned, we plan to adopt the method of ablation studies in the future. We plan to reduce the size of the training set and train multiple CGAN models while excluding different input channels. Then, we will compare the performance of these models to infer the contributions of the excluded channels.

We are exploring ways to improve our current model through new data inputs, alternative frameworks, and more advanced applications. First, we plan to train our model with data input from SSO satellites. The SSO satellites avoid the high zenith angle of geostationary satellites at high latitudes. Besides, imagers like VIIRS could provide nighttime data pairs of DNB and IR images for training as demonstrated in \cite{chirokova2023proxyvis, pasillas2024turning}. While \cite{han2022hypothetical} identifies the problem that their model has worse performance for areas with high diurnal temperature variations, training with nighttime data pairs would avoid this problem. This approach also allows us to include the MWIR band in the datasets, which was previously excluded because it is an emissive band at night but a reflective band in the day. The MWIR band has a much lower imaginary index of refraction for ice particles and water droplets than any of the IR bands currently adopted \cite{hale1973optical, warren2008optical}. This implies that the MWIR band can penetrate through much thicker cirrus and significantly improve the simulation of cirrus-covered areas. The significance of the MWIR band is also supported by \cite{chirokova2023proxyvis} which found that the MWIR band contributes the most to their ProxyVis model.

We are aware of the rapid advances in the field of DL image generation. The Diffusion models \cite{ho2020denoising, yang2023diffusion} have already surpassed the GAN models in various fields of image generation. Several Diffusion-based image-to-image translation frameworks such as Palette \cite{saharia2022palette} and Brownian Bridge Diffusion Models (BBDM) \cite{li2023bbdm} have been developed, some of which show promising applications in remote sensing \cite{xiao2023ediffsr, guo2024learning}. Given these advancements, we plan to replace our current pix2pix framework with BBDM to further enhance model performance, particularly for complex cloud features.

Beyond pixel-level simulations, we aim to incorporate broader meteorological metrics to enhance model validation and explore diverse model applications. Unlike current metrics such as SSIM and PSNR, object-based verification metrics can directly identify cloud patches and evaluate their similarity in real and AI-generated imagery from multiple aspects. For example, the Method for Object-based Diagnostic Evaluation (MODE) \cite{davis2006object} can be adapted for cloud reflectance and evaluate cloud simulations based on shape, position, and reflection intensity.

VIS imagery is widely used to derive meteorological information, including cloud optical depth and atmospheric motion vectors (AMVs). VIS reflectance plays a key role in retrieving cloud optical depth, which subsequently enables the derivation of cloud phase and cloud-top properties \cite{baum2000remote, baum2012modis}. The performance of our model can be further validated by substituting AI-generated reflectance into cloud property retrieval algorithms and comparing the results with those obtained using real VIS reflectance. Additionally, since cloud property retrieval is less accurate at night due to the absence of reflective bands, AI-generated reflectance could be combined with recent methods for simulating MWIR bands \cite{kim2019deep} to enhance nighttime retrieval accuracy. 

AMVs are estimations of winds derived from successive geostationary satellite images \cite{kazuki2017introduction}. The derivation of low-level AMVs relies on VIS images to track the low-level clouds accurately. When applied to model TC surface wind field, low-level AMVs derived from VIS images show lower RMSE, lower Bias, and improved coverage compared to the AMVs derived from IR images \cite{nonaka2019utilization}. With the capability to identify low clouds as VIS images do, our AI-generated images could be assimilated as satellite data to complement the derivation of nighttime low-level AMVs. An important application of low-level AMVs is to model the size of the TC wind field. With the AI-generated images as input, the TC wind field can be modeled accurately and continuously at night. Further studies are needed to confirm the feasibility of this potential application.

\section{Summary and conclusions}

VIS imagery has various important applications in meteorology, including monitoring TCs. However, it is unavailable at night because of the lack of sunlight. This study presents a CGAN model that simulates nighttime visible imagery with unprecedented accuracy and spatial resolution. Our model is a revised version of the pix2pix model in which we replaced the L1 loss with SSIM loss. We used AHI Band03 as the VIS band and carefully selected 7 IR bands (Band08, 09, 10, 11, 13, 15, 16), 4 direction parameters (solar zenith angle, solar azimuth angle, satellite zenith angle, satellite azimuth angle), and a land basemap as our model input. We obtained a total of 246078 daytime data pairs as our datasets. The statistical results of daytime model validation completely surpass the models presented in previous studies. This study also presents a nighttime model validation by comparing the AI-generated imagery to normalized VIIRS DNB radiance. Despite being slightly affected by the parallax, the nighttime model validation yields outstanding statistical results and demonstrates successful generalization to simulate nighttime cloud patterns. Additionally, we also highlighted a few potential applications of our model in TC monitoring. We showed that our model could be used to improve the positioning of relatively weak TC under vertical wind shear (VWS) and monitor the early stages of TC eye formation.

Our model successfully simulates visible satellite imagery with high accuracy from daytime to nighttime. The results of this study can significantly enhance nighttime monitoring of TCs and various other meteorological phenomena.

\section*{Acknowledgment}
We would like to thank Guihua Wang for very helpful discussions and suggestions. We thank Liangbo Qi, Lei Chen, and Xiaoqin Lu for their valuable comments on this article. We also thank Dike Su for support in the preliminary stages of the research and Fengsiyu Shu for help in finding TC cases. Additionally, the authors are grateful to anonymous reviewers for helpful and constructive comments. 

\bibliographystyle{IEEEtran}
\bibliography{IEEEabrv,reference}

\begin{thebibliography}{10}
\providecommand{\url}[1]{#1}
\csname url@samestyle\endcsname
\providecommand{\newblock}{\relax}
\providecommand{\bibinfo}[2]{#2}
\providecommand{\BIBentrySTDinterwordspacing}{\spaceskip=0pt\relax}
\providecommand{\BIBentryALTinterwordstretchfactor}{4}
\providecommand{\BIBentryALTinterwordspacing}{\spaceskip=\fontdimen2\font plus
\BIBentryALTinterwordstretchfactor\fontdimen3\font minus
  \fontdimen4\font\relax}
\providecommand{\BIBforeignlanguage}[2]{{%
\expandafter\ifx\csname l@#1\endcsname\relax
\typeout{** WARNING: IEEEtran.bst: No hyphenation pattern has been}%
\typeout{** loaded for the language `#1'. Using the pattern for}%
\typeout{** the default language instead.}%
\else
\language=\csname l@#1\endcsname
\fi
#2}}
\providecommand{\BIBdecl}{\relax}
\BIBdecl

\bibitem{bessho2016introduction}
K.~Bessho, K.~Date, M.~Hayashi, A.~Ikeda, T.~Imai, H.~Inoue, Y.~Kumagai,
  T.~Miyakawa, H.~Murata, T.~Ohno \emph{et~al.}, ``An introduction to
  himawari-8/9—japan’s new-generation geostationary meteorological
  satellites,'' \emph{Journal of the Meteorological Society of Japan. Ser. II},
  vol.~94, no.~2, pp. 151--183, 2016.

\bibitem{schmit2005introducing}
T.~J. Schmit, M.~M. Gunshor, W.~P. Menzel, J.~J. Gurka, J.~Li, and A.~S.
  Bachmeier, ``Introducing the next-generation advanced baseline imager on
  goes-r,'' \emph{Bulletin of the American Meteorological Society}, vol.~86,
  no.~8, pp. 1079--1096, 2005.

\bibitem{kim2021introduction}
D.~Kim, M.~Gu, T.-H. Oh, E.-K. Kim, and H.-J. Yang, ``Introduction of the
  advanced meteorological imager of geo-kompsat-2a: In-orbit tests and
  performance validation,'' \emph{Remote Sensing}, vol.~13, no.~7, p. 1303,
  2021.

\bibitem{yang2017introducing}
J.~Yang, Z.~Zhang, C.~Wei, F.~Lu, and Q.~Guo, ``Introducing the new generation
  of chinese geostationary weather satellites, fengyun-4,'' \emph{Bulletin of
  the American Meteorological Society}, vol.~98, no.~8, pp. 1637--1658, 2017.

\bibitem{conway1997introduction}
E.~D. Conway, \emph{An introduction to satellite image interpretation}.\hskip
  1em plus 0.5em minus 0.4em\relax JHU Press, 1997.

\bibitem{dvorak1984tropical}
V.~F. Dvorak, \emph{Tropical cyclone intensity analysis using satellite
  data}.\hskip 1em plus 0.5em minus 0.4em\relax US Department of Commerce,
  National Oceanic and Atmospheric Administration~…, 1984, vol.~11.

\bibitem{warren2008optical}
S.~G. Warren and R.~E. Brandt, ``Optical constants of ice from the ultraviolet
  to the microwave: A revised compilation,'' \emph{Journal of Geophysical
  Research: Atmospheres}, vol. 113, no. D14, 2008.

\bibitem{liao2013suomi}
L.~Liao, S.~Weiss, S.~Mills, and B.~Hauss, ``Suomi npp viirs day-night band
  on-orbit performance,'' \emph{Journal of Geophysical Research: Atmospheres},
  vol. 118, no.~22, pp. 12--705, 2013.

\bibitem{wang2020evaluation}
W.~Wang and C.~Cao, ``Evaluation of noaa-20 viirs reflective solar bands early
  on-orbit performance using daily deep convective clouds recent
  improvements,'' \emph{IEEE Journal of Selected Topics in Applied Earth
  Observations and Remote Sensing}, vol.~13, pp. 3975--3985, 2020.

\bibitem{kim2019nighttime}
K.~Kim, J.-H. Kim, Y.-J. Moon, E.~Park, G.~Shin, T.~Kim, Y.~Kim, and S.~Hong,
  ``Nighttime reflectance generation in the visible band of satellites,''
  \emph{Remote Sensing}, vol.~11, no.~18, p. 2087, 2019.

\bibitem{kim2020impact}
J.-H. Kim, S.~Ryu, J.~Jeong, D.~So, H.-J. Ban, and S.~Hong, ``Impact of
  satellite sounding data on virtual visible imagery generation using
  conditional generative adversarial network,'' \emph{IEEE Journal of Selected
  Topics in Applied Earth Observations and Remote Sensing}, vol.~13, pp.
  4532--4541, 2020.

\bibitem{harder2020nightvision}
P.~Harder, W.~Jones, R.~Lguensat, S.~Bouabid, J.~Fulton, D.~Quesada-Chac{\'o}n,
  A.~Marcolongo, S.~Stefanovi{\'c}, Y.~Rao, P.~Manshausen \emph{et~al.},
  ``Nightvision: generating nighttime satellite imagery from infra-red
  observations,'' \emph{arXiv preprint arXiv:2011.07017}, 2020.

\bibitem{han2022hypothetical}
K.-H. Han, J.-C. Jang, S.~Ryu, E.-H. Sohn, and S.~Hong, ``Hypothetical visible
  bands of advanced meteorological imager onboard the geostationary korea
  multi-purpose satellite-2a using data-to-data translation,'' \emph{IEEE
  Journal of Selected Topics in Applied Earth Observations and Remote Sensing},
  vol.~15, pp. 8378--8388, 2022.

\bibitem{cheng2022creating}
W.~Cheng, Q.~Li, Z.~Wang, W.~Zhang, and F.~Huang, ``Creating synthetic
  night-time visible-light meteorological satellite images using the gan
  method,'' \emph{Remote Sensing Letters}, vol.~13, no.~7, pp. 738--745, 2022.

\bibitem{yan2023simulation}
J.~Yan, J.~Qu, H.~An, and H.~Zhang, ``Simulation of visible light at night from
  infrared measurements using deep learning technique,'' \emph{Geocarto
  International}, no. just-accepted, pp. 1--13, 2023.

\bibitem{pasillas2024turning}
C.~M. Pasillas, C.~Kummerow, M.~Bell, and S.~D. Miller, ``Turning night into
  day: The creation and validation of synthetic nighttime visible imagery using
  the visible infrared imaging radiometer suite (viirs) day--night band (dnb)
  and machine learning,'' \emph{Artificial Intelligence for the Earth Systems},
  vol.~3, no.~3, p. e230002, 2024.

\bibitem{isola2017image}
P.~Isola, J.-Y. Zhu, T.~Zhou, and A.~A. Efros, ``Image-to-image translation
  with conditional adversarial networks,'' in \emph{Proceedings of the IEEE
  conference on computer vision and pattern recognition}, 2017, pp. 1125--1134.

\bibitem{chirokova2023proxyvis}
G.~Chirokova, J.~A. Knaff, M.~J. Brennan, R.~T. DeMaria, M.~Bozeman, S.~N.
  Stevenson, J.~L. Beven, E.~S. Blake, A.~Brammer, J.~W. Darlow \emph{et~al.},
  ``Proxyvis—a proxy for nighttime visible imagery applicable to
  geostationary satellite observations,'' \emph{Weather and Forecasting},
  vol.~38, no.~12, pp. 2527--2550, 2023.

\bibitem{alotaibi2020deep}
A.~Alotaibi, ``Deep generative adversarial networks for image-to-image
  translation: A review,'' \emph{Symmetry}, vol.~12, no.~10, p. 1705, 2020.

\bibitem{verhoef2012high}
A.~Verhoef, M.~Portabella, and A.~Stoffelen, ``High-resolution ascat
  scatterometer winds near the coast,'' \emph{IEEE Transactions on Geoscience
  and Remote Sensing}, vol.~50, no.~7, pp. 2481--2487, 2012.

\bibitem{kidder1995satellite}
S.~Q. Kidder and T.~H.~V. Haar, \emph{Satellite meteorology: an
  introduction}.\hskip 1em plus 0.5em minus 0.4em\relax Gulf Professional
  Publishing, 1995.

\bibitem{minnis1998parameterizations}
P.~Minnis, D.~P. Garber, D.~F. Young, R.~F. Arduini, and Y.~Takano,
  ``Parameterizations of reflectance and effective emittance for satellite
  remote sensing of cloud properties,'' \emph{Journal of the atmospheric
  sciences}, vol.~55, no.~22, pp. 3313--3339, 1998.

\bibitem{raffaelli1995cloud}
J.-L. Raffaelli and G.~M. Seze, ``Cloud type separation using local correlation
  between visible and infrared satellite images,'' in \emph{Passive Infrared
  Remote Sensing of Clouds and the Atmosphere III}, vol. 2578.\hskip 1em plus
  0.5em minus 0.4em\relax SPIE, 1995, pp. 61--67.

\bibitem{seze1987time}
G.~S{\`e}ze and W.~Rossow, ``Time-cumulated visible and infrared histograms
  used as descriptor of cloud cover,'' \emph{Advances in space research},
  vol.~7, no.~3, pp. 155--158, 1987.

\bibitem{jakel2013thermodynamic}
E.~J{\"a}kel, J.~Walter, and M.~Wendisch, ``Thermodynamic phase retrieval of
  convective clouds: impact of sensor viewing geometry and vertical
  distribution of cloud properties,'' \emph{Atmospheric Measurement
  Techniques}, vol.~6, no.~3, pp. 539--547, 2013.

\bibitem{menzel1997cloud}
W.~P. Menzel and K.~Strabala, \emph{Cloud top properties and cloud phase
  algorithm theoretical basis document}.\hskip 1em plus 0.5em minus 0.4em\relax
  University of Wisconsin--Madison, 1997.

\bibitem{menzel2008modis}
W.~P. Menzel, R.~A. Frey, H.~Zhang, D.~P. Wylie, C.~C. Moeller, R.~E. Holz,
  B.~Maddux, B.~A. Baum, K.~I. Strabala, and L.~E. Gumley, ``Modis global
  cloud-top pressure and amount estimation: Algorithm description and
  results,'' \emph{Journal of Applied Meteorology and Climatology}, vol.~47,
  no.~4, pp. 1175--1198, 2008.

\bibitem{lutz2003notes}
H.-J. Lutz, T.~Inoue, and J.~Schmetz, ``Notes and correspondence comparison of
  a split-window and a multi-spectral cloud classification for modis
  observations,'' \emph{Journal of the Meteorological Society of Japan. Ser.
  II}, vol.~81, no.~3, pp. 623--631, 2003.

\bibitem{gupta2022cloud}
R.~Gupta and S.~J. Nanda, ``Cloud detection in satellite images with classical
  and deep neural network approach: A review,'' \emph{Multimedia Tools and
  Applications}, vol.~81, no.~22, pp. 31\,847--31\,880, 2022.

\bibitem{tan2021detecting}
Z.~Tan, C.~Liu, S.~Ma, X.~Wang, J.~Shang, J.~Wang, W.~Ai, and W.~Yan,
  ``Detecting multilayer clouds from the geostationary advanced himawari imager
  using machine learning techniques,'' \emph{IEEE Transactions on Geoscience
  and Remote Sensing}, vol.~60, pp. 1--12, 2021.

\bibitem{goody1995atmospheric}
R.~M. Goody and Y.~L. Yung, \emph{Atmospheric radiation: theoretical
  basis}.\hskip 1em plus 0.5em minus 0.4em\relax Oxford university press, 1995.

\bibitem{Schmit2018ApplicationsOT}
T.~J. Schmit, S.~Lindstrom, J.~J. Gerth, and M.~M. Gunshor, ``Applications of
  the 16 spectral bands on the advanced baseline imager (abi).'' \emph{Journal
  of Operational Meteorology}, 2018.

\bibitem{saunders2018update}
R.~Saunders, J.~Hocking, E.~Turner, P.~Rayer, D.~Rundle, P.~Brunel, J.~Vidot,
  P.~Roquet, M.~Matricardi, A.~Geer \emph{et~al.}, ``An update on the rttov
  fast radiative transfer model (currently at version 12),''
  \emph{Geoscientific Model Development}, vol.~11, no.~7, pp. 2717--2737, 2018.

\bibitem{wu2020best}
Y.~Wu, F.~Zhang, K.~Wu, M.~Min, W.~Li, and R.~Liu, ``Best water vapor
  information layer of himawari-8-based water vapor bands over east asia,''
  \emph{Sensors}, vol.~20, no.~8, p. 2394, 2020.

\bibitem{hale1973optical}
G.~M. Hale and M.~R. Querry, ``Optical constants of water in the 200-nm to
  200-$\mu$m wavelength region,'' \emph{Applied optics}, vol.~12, no.~3, pp.
  555--563, 1973.

\bibitem{ellrod1995advances}
G.~P. Ellrod, ``Advances in the detection and analysis of fog at night using
  goes multispectral infrared imagery,'' \emph{Weather and Forecasting},
  vol.~10, no.~3, pp. 606--619, 1995.

\bibitem{calvert2010goes}
C.~Calvert and M.~Pavolonis, \emph{GOES-R advanced baseline imager (ABI)
  algorithm theoretical basis document for low cloud and fog}.\hskip 1em plus
  0.5em minus 0.4em\relax University of Wisconsin--Madison, Space Science and
  Engineering Center, 2010.

\bibitem{richter2005atmospheric}
R.~Richter and D.~Schl{\"a}pfer, ``Atmospheric/topographic correction for
  satellite imagery,'' \emph{DLR report DLR-IB}, vol. 565, 2005.

\bibitem{stockli2005blue}
R.~St{\"o}ckli, E.~Vermote, N.~Saleous, R.~Simmon, and D.~Herring, ``The blue
  marble next generation-a true color earth dataset including seasonal dynamics
  from modis,'' \emph{Published by the NASA Earth Observatory}, 2005.

\bibitem{ronneberger2015u}
O.~Ronneberger, P.~Fischer, and T.~Brox, ``U-net: Convolutional networks for
  biomedical image segmentation,'' in \emph{Medical Image Computing and
  Computer-Assisted Intervention--MICCAI 2015: 18th International Conference,
  Munich, Germany, October 5-9, 2015, Proceedings, Part III 18}.\hskip 1em plus
  0.5em minus 0.4em\relax Springer, 2015, pp. 234--241.

\bibitem{wang2004image}
Z.~Wang, A.~C. Bovik, H.~R. Sheikh, and E.~P. Simoncelli, ``Image quality
  assessment: from error visibility to structural similarity,'' \emph{IEEE
  transactions on image processing}, vol.~13, no.~4, pp. 600--612, 2004.

\bibitem{zinke2017simplified}
S.~Zinke, ``A simplified high and near-constant contrast approach for the
  display of viirs day/night band imagery,'' \emph{International Journal of
  Remote Sensing}, vol.~38, no.~19, pp. 5374--5387, 2017.

\bibitem{martin_raspaud_2023_10400258}
\BIBentryALTinterwordspacing
M.~Raspaud, D.~Hoese, P.~Lahtinen, G.~Holl, S.~Proud, S.~Finkensieper,
  A.~Dybbroe, A.~Meraner, J.~Strandgren, J.~Feltz, S.~Joro, X.~Zhang, BENR0,
  G.~Ghiggi, W.~Roberts, Youva, P.~de~Buyl, L.~Ørum Rasmussen, yukaribbba,
  mherbertson, J.~H.~B. Méndez, Y.~Zhu, rdaruwala, seenno, T.~Jasmin,
  BengtRydberg, Isotr0py, C.~Kliche, and T.~Barnie, ``pytroll/satpy: Version
  0.46.0 (2023/12/18),'' Dec. 2023. [Online]. Available:
  \url{https://doi.org/10.5281/zenodo.10400258}
\BIBentrySTDinterwordspacing

\bibitem{ohtake2013one}
M.~Ohtake, C.~Pieters, P.~Isaacson, S.~Besse, Y.~Yokota, T.~Matsunaga,
  J.~Boardman, S.~Yamomoto, J.~Haruyama, M.~Staid \emph{et~al.}, ``One moon,
  many measurements 3: Spectral reflectance,'' \emph{Icarus}, vol. 226, no.~1,
  pp. 364--374, 2013.

\bibitem{min2017investigation}
M.~Min, J.~Deng, C.~Liu, J.~Guo, N.~Lu, X.~Hu, L.~Chen, P.~Zhang, Q.~Lu, and
  L.~Wang, ``An investigation of the implications of lunar illumination
  spectral changes for day/night band-based cloud property retrieval due to
  lunar phase transition,'' \emph{Journal of Geophysical Research:
  Atmospheres}, vol. 122, no.~17, pp. 9233--9244, 2017.

\bibitem{uprety2017improving}
S.~Uprety, C.~Cao, Y.~Gu, and X.~Shao, ``Improving the low light radiance
  calibration of s-npp viirs day/night band in the noaa operations,'' in
  \emph{2017 IEEE International Geoscience and Remote Sensing Symposium
  (IGARSS)}.\hskip 1em plus 0.5em minus 0.4em\relax IEEE, 2017, pp. 4726--4729.

\bibitem{cao2019radiometric}
C.~Cao, Y.~Bai, W.~Wang, and T.~Choi, ``Radiometric inter-consistency of viirs
  dnb on suomi npp and noaa-20 from observations of reflected lunar lights over
  deep convective clouds,'' \emph{Remote Sensing}, vol.~11, no.~8, p. 934,
  2019.

\bibitem{chen2022two}
Y.-L. Chen and C.-C. Wu, ``On the two types of tropical cyclone eye formation:
  Clearing formation and banding formation,'' \emph{Monthly Weather Review},
  vol. 150, no.~6, pp. 1457--1473, 2022.

\bibitem{wang2021real}
X.~Wang, L.~Xie, C.~Dong, and Y.~Shan, ``Real-esrgan: Training real-world blind
  super-resolution with pure synthetic data,'' in \emph{Proceedings of the
  IEEE/CVF international conference on computer vision}, 2021, pp. 1905--1914.

\bibitem{lowe2004distinctive}
D.~G. Lowe, ``Distinctive image features from scale-invariant keypoints,''
  \emph{International journal of computer vision}, vol.~60, pp. 91--110, 2004.

\bibitem{liang2014improved}
C.~K. Liang, S.~Mills, B.~I. Hauss, and S.~D. Miller, ``Improved viirs
  day/night band imagery with near-constant contrast,'' \emph{IEEE Transactions
  on Geoscience and Remote Sensing}, vol.~52, no.~11, pp. 6964--6971, 2014.

\bibitem{ho2020denoising}
J.~Ho, A.~Jain, and P.~Abbeel, ``Denoising diffusion probabilistic models,''
  \emph{Advances in neural information processing systems}, vol.~33, pp.
  6840--6851, 2020.

\bibitem{yang2023diffusion}
L.~Yang, Z.~Zhang, Y.~Song, S.~Hong, R.~Xu, Y.~Zhao, W.~Zhang, B.~Cui, and
  M.-H. Yang, ``Diffusion models: A comprehensive survey of methods and
  applications,'' \emph{ACM Computing Surveys}, vol.~56, no.~4, pp. 1--39,
  2023.

\bibitem{saharia2022palette}
C.~Saharia, W.~Chan, H.~Chang, C.~Lee, J.~Ho, T.~Salimans, D.~Fleet, and
  M.~Norouzi, ``Palette: Image-to-image diffusion models,'' in \emph{ACM
  SIGGRAPH 2022 Conference Proceedings}, 2022, pp. 1--10.

\bibitem{li2023bbdm}
B.~Li, K.~Xue, B.~Liu, and Y.-K. Lai, ``Bbdm: Image-to-image translation with
  brownian bridge diffusion models,'' in \emph{Proceedings of the IEEE/CVF
  Conference on Computer Vision and Pattern Recognition}, 2023, pp. 1952--1961.

\bibitem{xiao2023ediffsr}
Y.~Xiao, Q.~Yuan, K.~Jiang, J.~He, X.~Jin, and L.~Zhang, ``Ediffsr: An
  efficient diffusion probabilistic model for remote sensing image
  super-resolution,'' \emph{IEEE Transactions on Geoscience and Remote
  Sensing}, 2023.

\bibitem{guo2024learning}
Z.~Guo, J.~Liu, Q.~Cai, Z.~Zhang, and S.~Mei, ``Learning sar-to-optical image
  translation via diffusion models with color memory,'' \emph{IEEE Journal of
  Selected Topics in Applied Earth Observations and Remote Sensing}, 2024.

\bibitem{davis2006object}
C.~Davis, B.~Brown, and R.~Bullock, ``Object-based verification of
  precipitation forecasts. part i: Methodology and application to mesoscale
  rain areas,'' \emph{Monthly Weather Review}, vol. 134, no.~7, pp. 1772--1784,
  2006.

\bibitem{baum2000remote}
B.~A. Baum, D.~P. Kratz, P.~Yang, S.~Ou, Y.~Hu, P.~F. Soulen, and S.-C. Tsay,
  ``Remote sensing of cloud properties using modis airborne simulator imagery
  during success: 1. data and models,'' \emph{Journal of Geophysical Research:
  Atmospheres}, vol. 105, no.~D9, pp. 11\,767--11\,780, 2000.

\bibitem{baum2012modis}
B.~A. Baum, W.~P. Menzel, R.~A. Frey, D.~C. Tobin, R.~E. Holz, S.~A. Ackerman,
  A.~K. Heidinger, and P.~Yang, ``Modis cloud-top property refinements for
  collection 6,'' \emph{Journal of applied meteorology and climatology},
  vol.~51, no.~6, pp. 1145--1163, 2012.

\bibitem{kim2019deep}
Y.~Kim and S.~Hong, ``Deep learning-generated nighttime reflectance and daytime
  radiance of the midwave infrared band of a geostationary satellite,''
  \emph{Remote Sensing}, vol.~11, no.~22, p. 2713, 2019.

\bibitem{kazuki2017introduction}
S.~Kazuki, ``Introduction to the himawari-8 atmospheric motion vector
  algorithm,'' 2017.

\bibitem{nonaka2019utilization}
K.~Nonaka, S.~Nishimura, and Y.~Igarashi, ``Utilization of estimated sea
  surface wind data based on himawari-8/9 low-level amvs for tropical cyclone
  analysis,'' \emph{RSMC Tokyo Typhoon Center Technical Review}, vol.~21, 2019.

\end{thebibliography}

\begin{IEEEbiography}[{\includegraphics[width=1in,height=1.25in,clip,keepaspectratio]{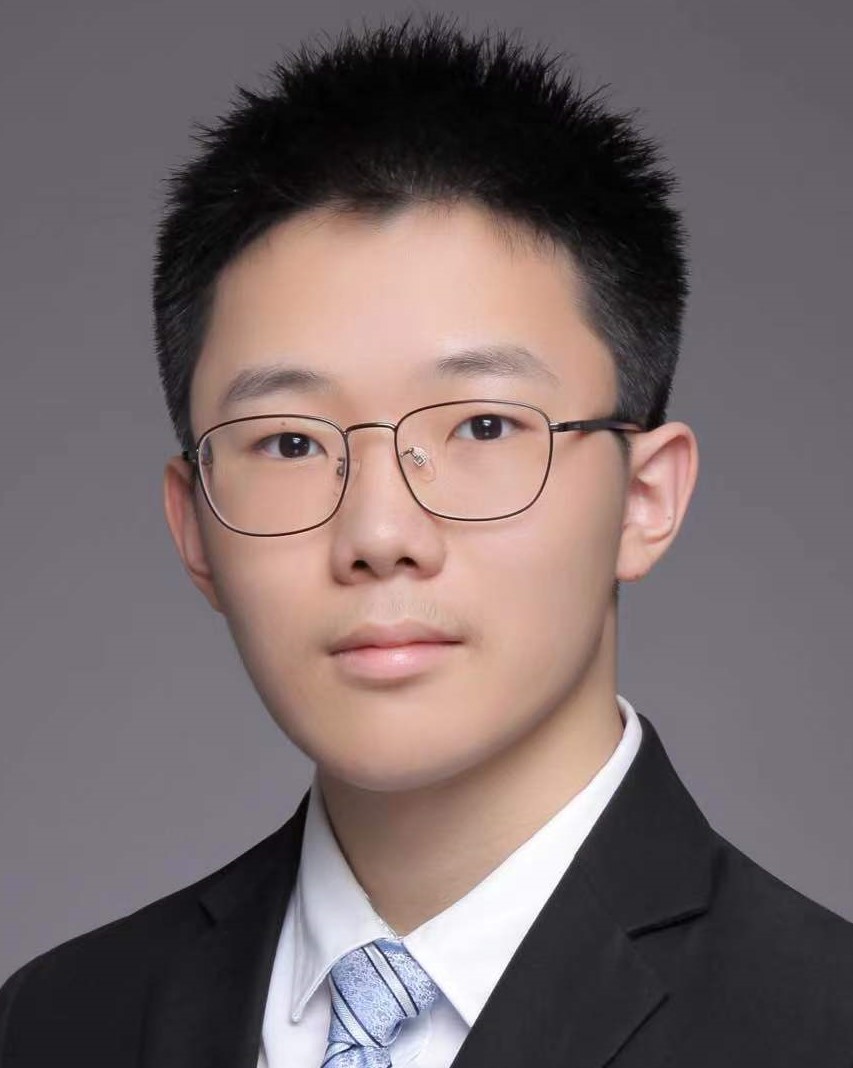}}]{Jinghuai Yao}
is currently an undergraduate student at the University of Wisconsin-Madison. His research interest includes image enhancement of satellite imagery based on both traditional and deep learning approaches, and applications of passive microwave imagers on tropical cyclones.
\end{IEEEbiography}

\begin{IEEEbiographynophoto}{Puyuan Du}
received the B.S. degree in chemistry from the University of California, Santa Barbara in 2023. He is currently working toward the M.S. degree in applied chemistry at the University of California, Los Angeles. His research interests include remote sensing data processing and interdisciplinary studies in chemistry and geoscience.
\end{IEEEbiographynophoto}

\begin{IEEEbiographynophoto}{Yucheng Zhao}
is currently an undergraduate student majoring in Information Resource Management at the School of Health Sciences, Guangzhou Xinhua University, Guangzhou, China. His research interests include remote sensing data processing and deep learning using remote sensing data.
\end{IEEEbiographynophoto}

\begin{IEEEbiographynophoto}{Yubo Wang}
is an undergraduate student at the University of Wisconsin-Madison majoring in Atmospheric and Oceanic Science (B.S.). His research interests include remote sensing applications on tropical cyclones based on both physical methods and artificial intelligence approaches, and artificial intelligence numerical weather prediction.
\end{IEEEbiographynophoto}

\end{document}